\documentclass[journal]{IEEEtran}

\usepackage{amsmath,graphicx}
\usepackage{algorithm}
\usepackage[noend]{algpseudocode}
\usepackage{stfloats}
\usepackage{xcolor}
\usepackage{amsfonts}
\usepackage{amssymb}
\usepackage{cite}
\usepackage{array}
\usepackage{subcaption}
\usepackage{times}
\usepackage{epsfig}
\usepackage{latexsym}
\usepackage{epstopdf}
\usepackage{verbatim}
\usepackage{units}
\usepackage{amsthm}
\usepackage{placeins}
\usepackage{afterpage}
\usepackage{dsfont}
\usepackage{soul}
\usepackage{multicol}
\usepackage{multirow}
\usepackage{mathtools}
\usepackage[cmintegrals]{newtxmath}
\usepackage{url}
\newcolumntype{P}[1]{>{\centering\arraybackslash}p{#1}}
\newcolumntype{M}[1]{>{\centering\arraybackslash}m{#1}}

\newcommand{\defeq}{\ensuremath{\triangleq}}

\ifCLASSINFOpdf

\else

\fi

\begin{document}
%
\title{Coded Distributed Computing with Partial Recovery}
%
%
%

\author{Emre~Ozfatura,
        Sennur~Ulukus,
        and~Deniz~G{\"u}nd{\"u}z
\thanks{This paper was presented in part at the 2019  IEEE International Conference on Acoustics, Speech and Signal Processing in Brighton, UK.}
\thanks{Emre Ozfatura and Deniz G{\"u}nd{\"u}z are with Information Processing and Communications Lab, Department of Electrical and Electronic Engineering,
Imperial College London Email: \{m.ozfatura, d.gunduz\} @imperial.ac.uk.}
\thanks{Sennur Ulukus is with the Department of Electrical and Computer Engineering, 
University of Maryland.} 
\thanks{This work was supported in part by the Marie Sklodowska-Curie Action SCAVENGE (grant agreement no. 675891), by the European Research Council (ERC) Starting Grant BEACON (grant agreement no. 677854), and UK EPSRC (EP/T023600/1) under the CHIST-ERA program (CHISTERA-18-SDCDN-001).}}


\maketitle

\begin{abstract}
Coded computation techniques provide robustness against \textit{straggling} workers in distributed computing. However, most of the existing schemes  require exact provisioning of the straggling behavior and ignore the computations carried out by straggling workers. Moreover, these schemes are typically designed to recover the desired computation results accurately, while in many machine learning and iterative optimization algorithms, faster approximate solutions are known to result in an improvement in the overall convergence time. In this paper, we first introduce a novel coded matrix-vector multiplication scheme, called \textit{coded computation with partial recovery (CCPR)}, which benefits from the advantages of both coded and uncoded computation schemes, and reduces both the computation time and the decoding complexity by allowing a trade-off between the accuracy
and the speed of computation. We then extend this approach to distributed implementation of more general computation tasks by proposing a coded communication scheme with partial recovery, where the results of subtasks computed by the workers are coded before being communicated. Numerical simulations on a large linear regression task confirm the benefits of the proposed scheme in terms of the trade-off between the computation accuracy and latency.  
\end{abstract}

\begin{IEEEkeywords}
Coded computation, distributed computation, maximum distance separable (MDS) code, linear codes, rateless codes, stragglers.  
\end{IEEEkeywords}

%
\IEEEpeerreviewmaketitle

\section{Introduction}
\label{sec:intro}
One of the key enablers of efficient machine learning solutions is the availability of large datasets. However, the ever growing size of the datasets and the complexity of the models trained on them lead also to an increase in the computational complexity and storage requirements of the algorithms employed. In parallel, there is a growing availability of cloud computing platforms (such as Amazon Web Services, Microsoft Azure and Google Cloud Functions) that offer computational resources to users to carry out demanding computation tasks. The associated distributed computation framework allows harnessing the computation and memory resources of multiple heterogeneous computation servers, referred to as {\em workers}.

In the most common implementation of distributed computation, a {\em parameter server (PS)} divides the main computational task into several subtasks and assigns them to workers.  Each worker executes the computation tasks assigned to it, and conveys the result to the PS. Having received the results from all the workers, the PS combines them to obtain the result of the main computation task. In principle, such a distributed computation framework should achieve a speed-up factor proportional to the number of workers employed. However, in real implementations, the overall computation time is constrained by the slowest, so-called {\em straggling worker(s)}. Moreover, as the number of employed workers increases, communication starts to become more complex and to introduce additional delays, which can aggravate the straggler problem. To remedy the delays due to straggling workers, various straggler-tolerant distributed computation schemes have been introduced recently, which build upon the idea of assigning redundant computations/subtasks to workers, to let faster workers  compensate for the stragglers \cite{UCUT.2,UCUT.3,UCUT.4,UCUT.5,UCUT.6,UCUT.7,UCUT.8,UCCT.1,UCCT.2,UCCT.3,UCCT.4,UCCT.5,UCCT.6,UCCT.7,CC.1,CC.2,CC.3,CC.4,CC.5,CC.6,CC.7,CC.8,CC.9,CC.11,CC.12,CC.13,CC.14,CC.15,CC.16,CC.17,CC.18,CC.19,CC.20,CC.21,CC.23,CC.31,CC.32,CC.33,CC.35,CC.AG1,CC.AG2,CC.AG3}.

\subsection{Motivation}
We will motivate the proposed distributed computation framework on a simple regression problem. In linear regression, the goal is to minimize the empirical mean squared-error:
\begin{equation}
L(\boldsymbol{\theta}) \triangleq \frac{1}{2N}\sum_{i=1}^{N}(y_{i}-\mathbf{x}_{i}^{T}\boldsymbol{\theta})^{2},
\label{loss}
\end{equation}
where $\mathbf{x}_{1},\ldots,\mathbf{x}_{N} \in \mathbb{R}^{d}$
are the data points with corresponding labels $y_{1},\ldots,y_{N} \in \mathbb{R}$, and $\boldsymbol{\theta}\in \mathbb{R}^{d}$ is the parameter vector. The optimal parameter vector can be obtained iteratively by gradient descent (GD), in which the parameter vector is updated iteratively as follows:
\begin{equation}
\boldsymbol{\theta}_{t+1}=\boldsymbol{\theta}_{t}-\eta_{t} \nabla_{\boldsymbol{\theta}} L(\boldsymbol{\theta}_{t}),
\end{equation}
 where $\eta_{t}$ is the learning rate at the $t$-th iteration. Gradient of the loss function in (\ref{loss}) can be written as 
 \begin{equation}\label{GD_eqn}
 \nabla_{\boldsymbol{\theta}} L(\boldsymbol{\theta}_t) = \mathbf{X}^{T} \mathbf{X} \boldsymbol{\theta}_{t}-\mathbf{X}^{T}\mathbf{y}, 
 \end{equation}
 where $\mathbf{X}=[\mathbf{x}_{1},\ldots,\mathbf{x}_{N}]^{T}$ and 
$\mathbf{y}=[y_{1},\ldots,y_{N}]^{T}$. In the gradient expression, only $\boldsymbol{\theta}_{t}$ changes over iterations; hence, the key computational task at each iteration is the  matrix-vector multiplication $\mathbf{W}\boldsymbol{\theta}_{t}$, where $\mathbf{W}\defeq\mathbf{X}^{T}\mathbf{X}\in\mathbb{R}^{d\times d}$.\\

\subsection{Coded Distributed Matrix-Vector Multiplication}
\indent The execution of $\mathbf{W}\boldsymbol{\theta}_{t}$ can be distributed across $K$ \textit{workers} by simply dividing $\mathbf{W}$ row-wise into $K$  disjoint submatrices and assigning each submatrix to one of the workers. However,
the computation time of this naive approach will  be limited by the {\em straggling} worker(s). The main challenge in this setup arises because the straggling behaviour (due either to the computation speed of the workers or the delays in communication) varies over time, and its realization at each iteration is not known in advance. On the other hand, statistical knowledge about the computation and communication latency for each worker can be acquired over time, and used for a more efficient allocation of computation tasks (e.g. as in \cite{CC.15,CC.23,CC.31}) as well as the coding scheme employed. For the sake of simplicity we assume homogeneous workers in this work.

{\em Coded computation} has been introduced to tolerate stragglers in matrix-vector multiplication by encoding the $\mathbf{W}$ matrix, and distributing the partitions of this encoded matrix among the workers, to achieve  redundancy \cite{CC.1,CC.2,CC.3,CC.4,CC.5,CC.6,CC.7,CC.8,CC.9,CC.11,CC.12,CC.13,CC.14}. One well-known method to introduce redundancy in matrix-vector multiplication  is to utilize  maximum distance separable (MDS) codes to encode $\mathbf{W}$ \cite{CC.1}. To elucidate the MDS-coded computation (MCC) we can divide $\mathbf{W}$ into $\bar{K}$ disjoint submatrices, $\mathbf{W}_{1},\ldots,\mathbf{W}_{\bar{K}}\in\mathbb{R}^{\bar{d}\times d}$, $\bar{d}=d/\bar{K}$,  which are then encoded with a $(\bar{K},K)$ MDS code. Each coded submatrix is assigned to a different worker, which  multiplies it with $\boldsymbol{\theta}_t$, and returns the result to the PS. The PS
 can recover $\mathbf{W}\boldsymbol{\theta}_{t}$ 
from the results of any $\bar{K}$ workers.\\
\indent Note that,
up to $K-\bar{K}$ stragglers can be tolerated with MCC at the expense of increasing the \textit{computation load} of each worker by $r = K/\bar{K}$; that is, each worker is assigned $r$ times more computations compared to the naive approach of equally dividing all the required computations among the workers. Alternative to MDS codes \cite{CC.1,CC.2,CC.8}, LDPC codes \cite{CC.3}, and rateless codes \cite{CC.9} have also been studied for straggler-tolerant coded computation in the literature.

\subsection{Computation-communication trade-off}
Conventional straggler-aware designs assume that a single message is transmitted by each worker after completing its assigned computation task. Under this limitation, straggler-aware schemes require exact provisioning of the straggler behaviour, and otherwise, suffer from {\em over-computation} and {\em under-utilization} \cite{overview}. To overcome these obstacles one can allow  each worker to send multiple messages to the PS at each iteration, which we refer to as {\em multi-message communication (MMC)} \cite{CC.2,CC.5,CC.9,CC.11,CC.33,UCUT.2,UCUT.4,UCCT.6}. However, MMC may introduce additional delays due to the communication overhead. Hence, with MMC the objective is to find an optimal operating point that balances the computation and communication latencies \cite{overview}. One recent approach for coded matrix-vector multiplication with MMC is to utilize rateless codes  \cite{CC.9} due to their advantages of  better utilizing the computational resources and low decoding complexity at the PS. However, in practice, rateless codes reach the target coding rates only if the number of coded messages are sufficiently large, which may not be desired in distributed computation framework since it leads to a congestion at the PS. Hence, the design of a code structure for distributed computation that can reduce the computation time without inducing an overwhelming communication overhead is an open challenge  that we address in this paper.

\subsection{Computation accuracy-computation speed trade-off}
In the case of iterative optimization algorithms we can consider the accuracy of the computations at each iteration as another dimension of the trade-off governing the overall convergence behaviour. For example, when applying gradient descent over large datasets, computation of the gradient in (\ref{GD_eqn}) at each iteration can become very costly. The most common alternative iterative optimization framework for large scale learning problems is  {\em stochastic gradient descent (SGD)}, which uses an estimate of the gradient in (\ref{GD_eqn}) at each iteration, evaluated on a random subset of the dataset. Hence, by changing the size of the sampled dataset it is possible to seek a balance between the accuracy of the gradient estimate and the computation time.  This trade-off is actually employed in several practical implementations\cite{adapt1, adapt2}.

On the other hand, a vast majority of  the coded computation schemes in the literature are designed for full gradient recovery. Nevertheless, a simple uncoded computation scheme with MMC \cite{UCUT.4,CC.5} can exploit partial computations performed by straggling workers, while also providing the PS a certain flexibility to terminate an iteration when a sufficient number of computations  are received. Accordingly, our goal here is to design a coded computing framework that can efficiently benefit from redundant computations with the flexibility of partial gradient computations. To this end, we introduce a novel hybrid scheme for distributed matrix-vector multiplication, called \textit{coded computation with partial recovery} (CCPR), bringing together the advantages of uncoded computation, such as low decoding complexity and partial gradient updates, with those of coded computation, such as reduced per-iteration completion time and reduced communication load.

We also want to highlight that, in most of the coded computation schemes in the literature, encoding is executed in PS in a centralized manner, and coded submatrices are distributed to workers; however, in our proposed strategy, as explained in Section \ref{RCS}, encoding step can be executed in a decentralized manner. Such local encoding provides two key advantages; first, it is possible to dynamically change the codewords over time based on the realization of the straggler behaviour \cite{CC.34}, which is particularly desired when the straggler behavior is correlated over time; second, as further explained in Section \ref{section:coded_comm}, it allows us to extend the CCPR strategy to more general distributed learning problems.

To the best of our knowledge, the partial recovery approach was  first introduced in our preliminary work \cite{CC.13}. In this paper, we extend this approach and provide a more comprehensive analysis. Our specific contributions can be summarized as follows:
\begin{itemize}
\item We provide a general framework, and highlight certain design principles to efficiently employ partial recovery in a coded computation scenario, particularly with MMC.
\item Based on these design principles, we introduce  {\em random circularly shifted (RCS) codes} for distributed matrix-vector multiplication.
\item We provide a generalization of RCS codes to the distributed implementation of computation tasks beyond matrix-vector multiplication by proposing a gradient coding scheme with partial computation.
\item Through numerical experiments in a linear regression problem, we show that RCS codes outperform existing distributed learning schemes across straggling workers, and also present the trade-offs between the update accuracy, communication latency and computation time achieved by these codes.
\end{itemize}

\begin{figure}
\centering
\includegraphics[scale=0.4]{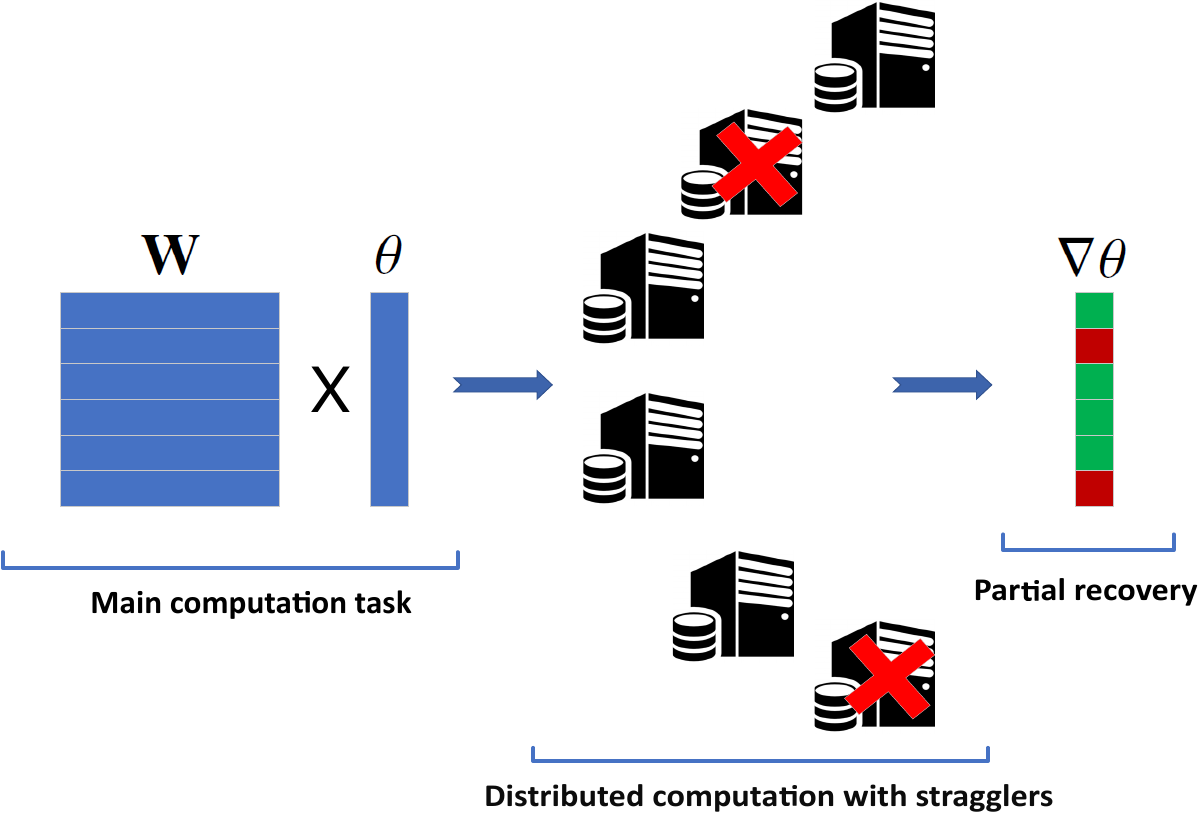}
\caption{Illustration of  partial recovery in a naive distributed computation scenario with 6 workers, 2 of which are stragglers.}
\label{partial_rec}
\end{figure}

\section{Coded computation with partial recovery}
In  conventional coded computation schemes, the PS waits until a sufficient number of computations are gathered from the workers to recover the correct results of the underlying computation task. In contrast, the partial recovery strategy does not necessarily aim at recovering the results accurately. In particular, for the matrix-vector multiplication task of $\mathbf{W}\boldsymbol{\theta}$, $\mathbf{W} \in \mathbb{R}^{d\times d}$, $\boldsymbol{\theta} \in \mathbb{R}^d$, we will target recovering only a subset of the entries of the $d$-dimensional result vector. We define the percentage of entries of $\mathbf{W}\boldsymbol{\theta}$ to be recovered as the {\em tolerance}, which will be dictated by the underlying computation task.\\
\indent For  encoding, we utilize a general linear code structure. Matrix $\mathbf{W}$ is initially divided into $K$ disjoint submatrices $\mathbf{W}_{1},\ldots,\mathbf{W}_{K}\in\mathbb{R}^{d/K\times d}$,  where we assume $K$ divides $d$, as otherwise this can be satisfied with zero padding. Then, $r$ coded submatrices, $\widetilde{\mathbf{W}}_{i,1},\ldots,\widetilde{\mathbf{W}}_{i,r}$, are assigned to each worker $i$ for computation, where each coded matrix $\widetilde{\mathbf{W}}_{i,j}$ is a linear combination of $K$ submatrices, i.e.,
\begin{equation}
\widetilde{\mathbf{W}}_{i,j}=\sum_{k\in[K]}\alpha^{(i)}_{j,k}\mathbf{W}_{k}.
\end{equation}
 Following the initial encoding phase, the $i$th worker performs the computations $\widetilde{\mathbf{W}}_{i,1}\boldsymbol{\theta},\ldots,\widetilde{\mathbf{W}}_{i,r}\boldsymbol{\theta}$ in the given order and sends the result as soon as it is completed. We remark that, in the considered MMC scenario, the order of the assigned coded computations  affects the completion time; therefore, we introduce the  {\em computation assignment matrix} $\mathbf{C}$ to represent a coded computation strategy, i.e,
 \[
\mathbf{C}\defeq
  \begin{bmatrix}
    \widetilde{\mathbf{W}}_{1,1} &  \widetilde{\mathbf{W}}_{2,1} &   \ldots &  \widetilde{\mathbf{W}}_{K,1}  \\
   \widetilde{\mathbf{W}}_{1,2} &  \widetilde{\mathbf{W}}_{2,2} &   \ldots &  \widetilde{\mathbf{W}}_{K,2}  \\
    \vdots &  \vdots &   \ldots &  \vdots  \\
     \widetilde{\mathbf{W}}_{1,r} &  \widetilde{\mathbf{W}}_{2,r} &   \ldots &  \widetilde{\mathbf{W}}_{K,r} 
  \end{bmatrix},
\]
where,  $\mathbf{C}$ indicates the execution order of the assigned computation tasks to each worker. More specifically,  submatrix $\mathbf{C}(j,i)$, $j \in [r]$, $i \in [K]$, denotes the $j$th computation task to be executed by the $i$th worker.

When  partial recovery is allowed, PS waits until $(1-q) \times 100$ percent of the entries of the result vector are successfully recovered. We call the parameter $q$ as the {\em tolerance}, which is a design parameter.

Consider a given a distributed coded computation strategy for $K$ workers and redundancy parameter $r$, denoted by $\Pi(K,r)$. Then, let $T_{\Pi(K,r)}(q)$ and $M_{\Pi(K,r)}(q)$ be the expected computation time and the expected number of received messages to recover $K(1-q)$ partial results, respectively. In the scope of this work, our goal is to design a coded computation strategy, for the scenario where the system operates at set of different tolerance levels $\mathcal{Q}$, and we want to achieve ''acceptable performance", both in terms of $T_{\Pi(K,r)}(q)$ and $M_{\Pi(K,r)}(q)$, within the whole spectrum of $q$. Based on system requirements and the computational task one can introduce certain constraints on these terms and can prioritize certain tolerance level. Hence, we want to introduce a novel coded computation framework so that one can search for the suitable coding strategy, utilizing the same structure, according to given performance requirements. On top of that proposed framework has two more advantages having a low decoding complexity and its re-configurable nature  which allows to update the coding strategy in parallel to a change in the performance requirements.


Let us first present a simple example to show how the proposed coded computation scheme with partial recovery can improve upon other schemes in the literature, such as MDS coding or uncoded computation with MMC (UC-MMC).

\subsection{Motivating example}
\label{subsec:mot}
Consider $K=4$ workers and assume that  $\mathbf{W}$ is divided into 4 submatrices $\mathbf{W}_1, \ldots, \mathbf{W}_4$. Let us first consider two known distributed computation schemes, namely UC-MM \cite{UCUT.4, CC.5} and MDS-coded computation (MCC) \cite{CC.1}. Each scheme is defined by its computation assignment matrix.\\
\indent In MDS-coded computation, linearly independent coded computation tasks are distributed to the workers as follows:
\[
\mathbf{C}_{MDS}=
\left[
  \begin{bmatrix}
    \mathbf{W}_{1}+ \mathbf{W}_{3}\\
    \mathbf{W}_{2}+\mathbf{W}_{4} 
  \end{bmatrix}
   \begin{bmatrix}
     \mathbf{W}_{1}+2\mathbf{W}_{3} \\
     \mathbf{W}_{2}+2\mathbf{W}_{4}  
  \end{bmatrix}
   \begin{bmatrix}
       \mathbf{W}_{1}+4\mathbf{W}_{3} \\
       \mathbf{W}_{2}+4\mathbf{W}_{4}  
  \end{bmatrix}
   \begin{bmatrix}
    \mathbf{W}_{1}+8\mathbf{W}_{3} \\
    \mathbf{W}_{2}+8\mathbf{W}_{4}  
  \end{bmatrix}
\right].
\]
$\mathbf{C}_{MDS}$ consists of a single row of computation tasks since each worker sends the results of its computations only after  all of them are completed,  e.g., first worker sends the concatenation of $[(\mathbf{W}_{1} + \mathbf{W}_{3})\boldsymbol{\theta} ~ \text{ }(\mathbf{W}_{2}+\mathbf{W}_{4})\boldsymbol{\theta}]$ after completing both computations. $\mathbf{C}_{MDS}$ above corresponds to a $(2,4)$ MDS code; hence, the PS can recover the full gradient from the results of any two workers.\\

\begin{table}[t]{\footnotesize
    \begin{center}
    \begin{tabular}{ | p{3.5cm} | p{0.8cm} | p{1.2cm} |p{0.8cm}|}
    \hline
    Cumulative computation type&   MCC $n_m(\mathbf{N}_{i})$ &  UC-MMC $n_u(\mathbf{N}_{i})$ & CCPR $n_c(\mathbf{N}_{i})$\\ \hline
    $\mathbf{N}_{1}:N_{2}=4,N_{1}=0,N_{0}=0$ &1 & 1 &1 \\ \hline $\mathbf{N}_{2}:N_{2}=3,N_{1}=1,N_{0}=0$ &4 &4 &4 \\ \hline $\mathbf{N}_{3}:N_{2}=3,N_{1}=0,N_{0}=1$ & 4& 4 &4 \\ \hline $\mathbf{N}_{4}:N_{2}=2,N_{1}=2,N_{0}=0$ &6&6 &6 \\ \hline $\mathbf{N}_{5}:N_{2}=2,N_{1}=1,N_{0}=1$ & 12& 8 &12 \\ \hline $\mathbf{N}_{6}:N_{2}=2,N_{1}=0,N_{0}=2$ & 6&2 &6 \\ \hline
  $\mathbf{N}_{7}:N_{2}=1,N_{1}=3,N_{0}=0$ & 0&4 &4 \\ \hline $\mathbf{N}_{8}:N_{2}=1,N_{1}=2,N_{0}=1$ & 0& 4 &8 \\ \hline
$\mathbf{N}_{9}:N_{2}=0,N_{1}=4,N_{0}=0$ & 0& 1 &1 \\ \hline	
		\end{tabular}
		\caption{Number of  successful score vectors for each cumulative computation that can accurately recover the computation task with $K=4$ and $r=2$.}\label{table: ex1full}	          
		\end{center}
		       }
		       \vspace*{-0.8cm}
\end{table}

\indent In the UC-MMC scheme with cyclic shifted computation assignment \cite{CC.5}, the computation scheduling matrix is given by
\[
\mathbf{C}_{UC-MMC}=
  \begin{bmatrix}
    \mathbf{W}_{1} & \mathbf{W}_{2} & \mathbf{W}_{3} & \mathbf{W}_{4} \\
   \mathbf{W}_{2} & \mathbf{W}_{3}& \mathbf{W}_{4}  & \mathbf{W}_{1} \\
  \end{bmatrix},
\]
and each worker sends the results of its computations sequentially, as soon as each of them is completed. This helps to reduce the per-iteration completion time with an increase in the communication load \cite{UCUT.4, CC.5}.  With UC-MMC, full gradient can be recovered even if each worker performs only one computation, which is faster if the workers have similar speeds.\\
\indent The computation scheduling matrix of the proposed CCPR scheme is given by
 \[
\mathbf{C}_{CCPR}=
  \begin{bmatrix}
    \mathbf{W}_{1} & \mathbf{W}_{2} & \mathbf{W}_{3} & \mathbf{W}_{4} \\
    \mathbf{W}_{3}+\mathbf{W}_{4}& \mathbf{W}_{1}+\mathbf{W}_{3} & \mathbf{W}_{2}+\mathbf{W}_{4} & \mathbf{W}_{1}+\mathbf{W}_{2}  \\  
  \end{bmatrix},
  \]
which is a combination of the uncoded and coded approaches. Below we illustrate the advantages of this scheme by comparing its performance for both accurate and approximate computations. For this analysis we need to introduce a few definitions.\\
\indent Let $N_{s}(t)$ denote the number of workers that have completed exactly $s$  computations by time $t$, $s = 0, \ldots, r$. We define  $\mathbf{N}(t) \triangleq (N_{r}(t),\ldots,N_{0}(t))$ as the {\em cumulative computation type} at time $t$. Additionally, we introduce the $K$-dimensional \textit{score vector} $\mathbf{S}(t)=[s_{1}(t),\ldots,s_{K}(t)]$, where $s_{i}(t)$ denotes the number of computations completed and communicated by the $i$th worker by time $t$. We will call a score vector {\em successful} if it allows the recovery of the desired computation task at the PS. We note that, due to the homogeneous worker assumption, the probability of experiencing any score vector with the same cumulative computation type is the same. Therefore, what is important for the overall computation time statistics is the number of successful score vectors corresponding to each computation type.

\subsection{Full gradient performance}

\indent Let $n_m(\mathbf{N}), n_u(\mathbf{N})$ and $n_c(\mathbf{N})$ denote the number of distinct succesfull score vectors with the cumulative computation type $\mathbf{N}$ that allow the recovery of $\mathbf{W}\boldsymbol{\theta}$ for the MDS, UC-MMC, and CCPR schemes, respectively. For instance, for the cumulative computation type $\mathbf{N}=(1,2,1)$, MDS scheme cannot recover the full gradient; however, UC-MMC can recover the full gradient for four $\mathbf{S}$ vectors; $\mathbf{S}=[2,0,1,1]$, $\mathbf{S}=[1,2,0,1]$, $\mathbf{S}=[1,1,2,0]$ and $\mathbf{S}=[0,1,1,2]$, hence $n_u(\mathbf{N})=4$. Finally, in the CCPR scheme, there are in total $n_c(\mathbf{N}) =8$ successful $\mathbf{S}$ vectors; $[2,1,0,1]$, $[2,1,1,0]$, $[1,2,0,1]$, $[0,2,1,1]$, $[1,0,2,1]$, $[1,1,2,0]$, $[0,1,1,2]$ and $[1,0,1,2]$.\\
\indent These values are listed in Table \ref{table: ex1full} for the cumulative computation types with at least one successful score vector for one of the schemes.
Particularly striking are the last three rows that correspond to cases with very few computations completed, i.e., when at most one worker completes all its assigned tasks. In these cases, CCPR is much more likely to recover $\mathbf{W}\boldsymbol{\theta}$; and hence, the computation deadline can be reduced significantly.
\begin{table}[t]{\footnotesize
    \begin{center}
    \begin{tabular}{ | p{3.5cm} | p{1.2cm} | p{1.2cm} | p{1.2cm} |}
    \hline
    Cumulative computation type &  MCC $n_m(\mathbf{N}_{i})$ &  UC-MM $n_u(\mathbf{N}_{i})$ & CCPR $n_c(\mathbf{N}_{i})$\\ \hline
    $\mathbf{N}_{1}:N_{2}=4,N_{1}=0,N_{0}=0$ &1 &1 &1 \\ \hline $\mathbf{N}_{2}:N_{2}=3,N_{1}=1,N_{0}=0$ &4 &4 &4 \\ \hline $\mathbf{N}_{3}:N_{2}=3,N_{1}=0,N_{0}=1$ &4 &4 &4 \\ \hline $\mathbf{N}_{4}:N_{2}=2,N_{1}=2,N_{0}=0$ &6 &6 &6 \\ \hline $\mathbf{N}_{5}:N_{2}=2,N_{1}=1,N_{0}=1$ &12 &12 &12 \\ \hline $\mathbf{N}_{6}:N_{2}=2,N_{1}=0,N_{0}=2$ &6 &6 &6 \\ \hline
  $\mathbf{N}_{7}:N_{2}=1,N_{1}=3,N_{0}=0$ &0 &4 &4 \\ \hline $\mathbf{N}_{8}:N_{2}=1,N_{1}=2,N_{0}=1$ &0 &12 &12 \\ \hline
 $\mathbf{N}_{9}:N_{2}=1,N_{1}=1,N_{0}=2$ &0 &8 &8 \\ \hline
 $\mathbf{N}_{10}:N_{2}=0,N_{1}=4,N_{0}=0$ &0 & 1 &1 \\ \hline
 $\mathbf{N}_{11}:N_{2}=0,N_{1}=3,N_{0}=1$ &0 & 4 &4 \\ \hline
		\end{tabular}
		\caption{Number of  successful score vectors for
		each cumulative computation type that can result in the recovery of at least 3 out of 4 computations with $K=4$ and $r=2$.}\label{table: ex1par}	          
		\end{center}
		       }
		       \vspace*{-0.8cm}
\end{table}
For a more explicit comparison of the completion time statistics, we can analyze the  probability of each type under a specific computation time statistics. Then, the probability of cumulative computation type $\mathbf{N}=(N_r,\ldots,N_0)$ at time $t$ is given by
\begin{equation}
\mathrm{Pr}(\mathbf{N}(t)=\mathbf{N})=\prod_{s=0}^{r} P_{s}(t)^{N_{s}},
\end{equation}
where  $P_{s}(t)$ is the probability of completing exactly $s$ computations by time $t$. Let $T$ denote the recovery time of the desired computation. Accordingly, for any of the schemes, we can write $\mathrm{Pr}(T<t)=\sum_{i=1}^9 n_a(\mathbf{N}_{i}) \cdot \mathrm{Pr}(\mathbf{N}(t)=\mathbf{N}_i)$, $a\in\{m,u,c\}$, where the types  $\mathbf{N}_{i}$ and corresponding $n_a(\mathbf{N}_{i})$, $i=1, \ldots, 9$, are listed in Table \ref{table: ex1full}. It is now clear that CCPR has the highest $\mathrm{Pr}(T<t)$ for any $t$; and hence, the minimum average completion time $E[T]$.  In the next subsection, we will highlight the partial recovery property of CCPR.    

\subsection{Partial computation performance}
Now, we compare the three schemes when we reduce the tolerance level, and aim at recovering only a portion of the computation results. In particular, for the above example, we will deem a scheme successful if it recovers at least 3 out of 4 values, $\left\{\mathbf{W}_{1}\boldsymbol{\theta},\ldots,\mathbf{W}_{4}\boldsymbol{\theta}\right\}$, corresponding to a tolerance of $25\%$. For each cumulative computation type the number of successful score vectors are listed in Table \ref{table: ex1par}. We can see that UC-MMC and CCPR have the same average completion time statistics. Hence, CCPR can provide a lower average per-iteration completion time for accurate computation compared to UC-MMC, while achieving  the same performance when partial computation is allowed.
\section{Design Principles of CCPR}\label{sec:Design}
For the encoding of the assigned computations we use a similar strategy to {\em rateless codes}, particularly to LT codes \cite{lt}. We first briefly explain the LT code structure, and highlight the required modification for our problem setup.\\
\indent Consider a sequence of symbols $\overline{\mathbf{W}}=\left\{\mathbf{W}_{1},\ldots,\mathbf{W}_{K}\right\}$ (in our setup these correspond to submatrices $\mathbf{W}_{i}$) to be transmitted over an erasure channel which correspond to stragglers in our model. The codewords (coded computations in our model) are formed as linear combinations of $\mathbf{W}_{1},\ldots,\mathbf{W}_{K}$, and the goal is to correctly recover the original sequence from only a random subset of the coded symbols. In the encoding phase, a coded symbol is formed by choosing $d$ elements randomly from $\overline{\mathbf{W}}$ and summing them, where $d$, which simply defines the degree of the symbol, comes from a distribution $P(d)$. In the decoding part, each coded symbol is decomposed by using  the recovered symbols, that is if a coded symbol contains a previously recovered symbol then it is subtracted from the coded symbol to obtain a new coded symbol with a smaller degree. Overall the objective is to recover all $K$ symbols from $K(1+\epsilon)$ coded symbols with  as small an $\epsilon$ as possible which reflects the overhead.\\
\indent We remark that coded symbols with smaller degrees can be decomposed faster; however, having many coded symbols with smaller degrees increases the probability of linear dependence among codewords. Hence, the degree distribution plays an important role in the performance of  LT codes. It has been shown that for a carefully chosen  $P(d)$,  $\epsilon$ goes to zero as $K\rightarrow\infty$. LT codes have the following drawbacks when employed in distributed computation. 

\begin{algorithm}[t]
\caption{RCS coded computation}\label{alg:dist_comp}
\begin{algorithmic}[1]
\State{\textbf{Data assignment phase:}}
\State{$L=\sum^{m}_{i=1}d_{i}$}
\State{Choose  random subset $\mathcal{I}\subset[K]$,  $\vert\mathcal{I}\vert=L$}
\For{row index $i=1,\ldots,L$}
\State{Randomly choose  $j\in\mathcal{I}$}
\State{Update $\mathcal{I}$: $\mathcal{I} \leftarrow \mathcal{I} \setminus \left\{ j\right\}$}
\State{$\mathbf{A}(i,:) = circshift(\overline{\mathbf{W}},j-1)$}
\EndFor
\State{\textbf{Data encoding phase:}}
\For{worker $k=1,\ldots,K$}
\For{message $j=1,\ldots,r$}
\State{Starting row index: $l_s = \sum^{j-1}_{i=1}d_i+1$}
\State{Ending row index: $l_e = \sum^{j}_{i=1}d_i$}
\State{$\widetilde{\mathbf{W}}_{i,j}=\sum^{l_e}_{l=l_s} \mathbf{A}(l,i)$}
\EndFor
\EndFor
\end{algorithmic}
\end{algorithm}

First,  the LT codes are designed under the assumption of  receiving a large number of coded symbols; however, in a distributed computation scenario the number of symbols is typically limited since each symbol corresponds to a message transmitted to the PS over the network, and increasing the number of messages may lead to congestion and communication delays at some point. On the other hand, for small $K$ the overhead $\epsilon$ might be high.\\ 
\indent Second, the degree distribution of LT codes, $P(d)$,
is designed for the correct recovery of the original sequence $\overline{\mathbf{W}}$. However, in our partial coded computation scenario we want to provide a certain flexibility by allowing the recovery of only $(1-q)K$ symbols, for some predefined tolerance $q$. Although the partial recovery of the rateless codes has been studied in the literature \cite{partrec}, we note that partial recovery for coded computation requires a tailored approach since the computational tasks, each of which corresponding to a distinct coded symbol, are executed sequentially; thus, the corresponding erasure probabilities due to straggling behavior of workers, are neither identical nor independent. Therefore, the coded symbols must be designed taking into account their execution orders to prevent overlaps and to minimize the average completion time.\\ 
\indent Hence, the main idea behind the CCPR scheme is to utilize the LT code structure in a more systematic way such that the degree of the each coded computation task is chosen carefully based on its computation order and with aim of  partial recovery.

\subsection{Computation order and degree limitation}
 We want to re-emphasize that, coded computation tasks with lower degrees can be recovered faster but as the number of computations received at the PS increases lower degree computations become less and less informative. Therefore, we want initial computations to be easily recoverable while those completed later to be more informative. Hence, we introduce the following design criteria; (i) for the first row of the computation assignment matrix, we consider uncoded computations, (ii) for a particular worker and  computation orders $i<j<r$, the degree of the computation at order $i$ can not be higher than that of the one at order $j$.
 
\subsection{Uniformity imposed encoding}
As highlighted before, coded messages with lower degrees may result in duplicate recoveries, wasting the computation resources. To this end, under the specified degree limitation, the main design challenge is how to form the coded computations  to prevent duplicate messages as much as possible. Accordingly, the challenge is to distribute submatrices $\mathbf{W}_{1},\ldots,\mathbf{W}_{K}$ among the coded computation tasks in a uniform fashion. We introduce two types of uniformity to help reduce the complexity of code design by introducing some structure to the code.

\subsubsection{Order-wise uniformity}
By  order-wise uniformity, we impose a constraint on the code construction such that  computations with the same order must have the same degree, and among the computations with the same  order, each submatrix $\mathbf{W}_{k}$ must appear in exactly the same number of computations. Formally speaking, let
\begin{equation}
\widetilde{\mathcal{W}}^{j}_{k}\defeq\left\{ \widetilde{\mathbf{W}}_{i,j}:
\alpha^{i}_{j,k}\neq0, \text{ }i\in[K]\right\}
\end{equation}
be the set of coded computations at order $j\in[r]$ containing submatrix $\mathbf{W}_{k}$. Then, the order-wise uniformity constraint imposes
\begin{equation}
\vert\widetilde{\mathcal{W}}^{j}_{k}\vert = d_{j}, \text{ }\forall j\in[r], \text{ } \forall k\in[K],
\end{equation}
for some $d_{j}$.

\subsubsection{Worker-wise uniformity}
Worker-wise uniformity imposes a constraint on the coded computations assigned to each worker, such that the coded computations do not contain any common submatrices. Formally speaking, for any worker $i\in[K]$, if $\alpha^{i}_{j,k} \neq 0$, for some $j\in[r]$  and $k\in[K]$, then $\alpha^{i}_{l,k}=0$, $\forall~ l\in[r]\setminus\left\{j\right\}$.\\
\indent Next, we introduce an encoding structure which ensures both order-wise and worker-wise uniformity.

\section{Randomly Circular Shifted (RCS) Code Design}\label{RCS}

In this section, we  introduce the randomly circular shifted (RCS) code design for coded distributed computation, which consists of two steps; namely data assignment and code construction. Before explaining these steps in detail, we first define the degree vector $\mathbf{d}$ of length $r$, whose $i$th entry $d_i$ denotes the degree of the  coded computations assigned to workers in order $i \in [r]$. Based on the aforementioned design criteria we set $d_1=1$ and,  for any  $i<j$, we have   $d_{i} \leq d_{j}$. Once the degree vector $\mathbf{d}$ is fixed, the two phases of RCS code design can be implemented.

In the first phase, a data assignment matrix is formed by using random circular shifts on the vector of submatrices $\overline{\mathbf{W}}=[\mathbf{W}_{1},\ldots,\mathbf{W}_{K}]$. At the beginning of the first phase, an index set $\mathcal{I}\subset[K]$ of size $L=\sum^{r}_{i=1}d_i$ is randomly chosen, where $L$ here denotes the total number of partial computations that can be recovered from a worker. Then using the elements of $\mathcal{I}$ as a parameter of the circular shift operator on vector $\overline{\mathbf{W}}$, a data assignment matrix $\mathbf{A}_{RCS}$ is formed following Algorithm 1 (line 5-8).

Let us illustrate the RCS code on a simple example. Consider $K=20$ workers and $\mathbf{d} = [1 2 3]$. This means $L=6$, and assume that $\mathcal{I} = \left\{1, 4, 11, 15, 6, 18 \right\}$. Then, for the $i$th row of $\mathbf{A}_{RCS}$,  $j\in\mathcal{I}$, is chosen randomly and discarded from $\mathcal{I}$, while  $\overline{\mathbf{W}}$ is circularly shifted by $j-1$. For the sake of simplicity, we assume that elements of $\mathcal{I}$ are chosen with the given order $1,4,11,15,6,18$, and the corresponding data assignment matrix is illustrated in Fig. \ref{assignment}.

\begin{figure}
 \[
\mathbf{A}_{RCS}=
  \begin{bmatrix}
  {\color{blue}\mathbf{W}_{1}} & {\color{blue}\mathbf{W}_{2}} & {\color{blue}\mathbf{W}_{3}} & \dots  & {\color{blue}\mathbf{W}_{20}} \\
  {\color{red}\mathbf{W}_{4}}\ & {\color{red}\mathbf{W}_{5}} & {\color{red}\mathbf{W}_{6}} & \dots  & {\color{red}\mathbf{W}_{3}} \\
  {\color{red}\mathbf{W}_{11}} & {\color{red}\mathbf{W}_{12}} & {\color{red}\mathbf{W}_{13}} & \dots  & {\color{red}\mathbf{W}_{10}}\\
  {\color{brown}\mathbf{W}_{15}} & {\color{brown}\mathbf{W}_{16}} & {\color{brown}\mathbf{W}_{17}} & \dots  & {\color{brown}\mathbf{W}_{14}} \\
  {\color{brown}\mathbf{W}_{6}} & {\color{brown}\mathbf{W}_{7}} & {\color{brown}\mathbf{W}_{8}} & \dots  & {\color{brown}\mathbf{W}_{5}} \\
  {\color{brown}\mathbf{W}_{18}} & {\color{brown}\mathbf{W}_{19}} & {\color{brown}\mathbf{W}_{20}} & \dots  & {\color{brown}\mathbf{W}_{17}}\\
  \end{bmatrix}\label{rcsc_dist}
  \]
\caption{Data assignment matrix for $K=20$ and $\mathcal{I}=\left\{1,4,11,15,6,18\right\}$.}
\label{assignment}
\end{figure}

Once the data assignment matrix is fixed, codewords  can be generated based on the degree vector $\mathbf{d}$ for each column independently and identically. The colors in the data assignment matrix in Fig. \ref{assignment} represent the submatrices that will form the same coded computation. The computation assignment for the first user, using the submatrices on the first column of the data assignment matrix, is illustrated in Fig. \ref{rcsc}.

We note that each coded message  corresponds to a linear equation, and in the decoding phase any approach for solving a set of  linear equations can be utilized, e.g., we can form a matrix with the coefficients of the coded messages, $\alpha^{i}_{j,k}$, that is, each coded message is represented by a binary row vector, and obtain the {\em reduced row echelon} form. Similarly to LT codes, we consider a low complexity decoding framework that decomposes the codewords successively by using only recovered symbols. From the construction, a codeword of degree $d$ is reduced $d-1$ time to recover an input symbol. Also, since the first $K$ codewords are of degree one, the total number of reduction on the codewords is limited by $K \times \tilde{L}$ where $\tilde{L}=\sum^{r}_{i=2}d_i-1$; and hence, the worst case  complexity of the decoding phase is $\mathcal{O}(K \tilde{L})$.

\begin{figure}
\[
  \begin{bmatrix}
  {\color{blue}\mathbf{W}_{1}} \\
  {\color{red}\mathbf{W}_{4}} \\
  {\color{red}\mathbf{W}_{11}} \\
  {\color{brown}\mathbf{W}_{15}} \\
  {\color{brown}\mathbf{W}_{6}} \\
  {\color{brown}\mathbf{W}_{18}} \\
  \end{bmatrix} \rightarrow \mathbf{C}_{1}=
  \begin{bmatrix}
   \widetilde{\mathbf{W}}_{1,1} \\
   \widetilde{\mathbf{W}}_{1,2}  \\
  \widetilde{\mathbf{W}}_{1,3} \\
  \end{bmatrix} =
  \begin{bmatrix}
   \mathbf{W}_{1} \\
   \mathbf{W}_{4} + \mathbf{W}_{11} \\
   \mathbf{W}_{15} + \mathbf{W}_{6} + \mathbf{W}_{18}\\
  \end{bmatrix}
 \]
  \caption{Illustration of the encoding phase for the first worker and the corresponding first column of the computation assignment matrix, $\mathbf{C}_{1}$.}
  \label{rcsc}
\end{figure}
\section{Extension to coded communication
scenario}\label{section:coded_comm}
Above, we have mainly focused on coded computation in the context of a linear regression problem, where the main computation task boils down to distributed matrix-vector multiplication. In a more general distributed computation problem, in which the computations cannot be expressed as a linear transform of the dataset, we cannot employ a similar coded computation technique. However, if the overall computation task can be written as the summation of smaller partial computation tasks, then, the redundancy can be achieved by assigning each of these partial computations to multiple workers. Communication load of such an implementation can be reduced by coded communication, where each worker sends to PS linear combinations of its partial computations. The gradient coding (GC) scheme, introduced in \cite{UCCT.1}, considers gradient estimates on subsets of a dataset as partial computations, and achieves redundancy by replicating parts of the dataset at multiple workers. This approach has been extended in various directions to improve the performance \cite{UCCT.2,UCCT.3,UCCT.4,UCCT.5,UCCT.6,UCCT.7}.\\
 \indent Let $\mathcal{G}=\{\mathbf{g}_{1},\ldots,\mathbf{g}_{K}\}$ be the results corresponding to $K$ partial computations, and the goal is to recover their sum at PS. In the GC scheme with computation load $r$, $r$ partial computations, denoted by $\mathcal{G}_{k}$, are assigned to worker $k$. Each worker, after completing $r$ partial computations, sends a linear combination of its results to the PS
\begin{equation}
\mathbf{c}_{k}\defeq\mathcal{L}_{k}(\mathbf{g}_{i}:\mathbf{g}_{i}\in\mathcal{G}_{k}).
\end{equation}

We refer to the linear combinations $\mathbf{c}_{1},\ldots,\mathbf{c}_{K}$ as {\em coded  partial computations}. The PS waits until it receives sufficiently many coded partial computations to recover the full gradient. It is shown in \cite{UCCT.1} that, for any set of non-straggler workers $\tilde{\mathcal{K}}\subseteq[K]$ with  $ \lvert \tilde{\mathcal{K}}  \rvert = K-r+1 $, there exists a set of coefficients $\mathcal{A}_{\tilde{\mathcal{K}}}=\left \{a_{k}:k\in\tilde{\mathcal{K}}\right\}$ such that
\begin{equation}
\sum_{k\in\tilde{\mathcal{K}}} a_{k}c^{(t)}_{k}=\sum^{K}_{k=1}g^{(t)}_{k}.
\end{equation} 
\begin{figure}
 \[
\mathbf{A}_{RCS}=
  \begin{bmatrix}
  {\color{blue}\mathbf{g}_{1}} & {\color{blue}\mathbf{g}_{2}} & {\color{blue}\mathbf{g}_{3}} & \dots  & {\color{blue}\mathbf{g}_{20}} \\
  {\color{red}\mathbf{g}_{4}}\ & {\color{red}\mathbf{g}_{5}} & {\color{red}\mathbf{g}_{6}} & \dots  & {\color{red}\mathbf{g}_{3}} \\
  {\color{red}\mathbf{g}_{11}} & {\color{red}\mathbf{g}_{12}} & {\color{red}\mathbf{g}_{13}} & \dots  & {\color{red}\mathbf{g}_{10}}\\
  {\color{brown}\mathbf{g}_{15}} & {\color{brown}\mathbf{g}_{16}} & {\color{brown}\mathbf{g}_{17}} & \dots  & {\color{brown}\mathbf{g}_{14}} \\
  {\color{brown}\mathbf{g}_{6}} & {\color{brown}\mathbf{g}_{7}} & {\color{brown}\mathbf{g}_{8}} & \dots  & {\color{brown}\mathbf{g}_{5}} \\
  {\color{brown}\mathbf{g}_{18}} & {\color{brown}\mathbf{g}_{19}} & {\color{brown}\mathbf{g}_{20}} & \dots  & {\color{brown}\mathbf{g}_{17}}\\
  \end{bmatrix}.
  \]
\caption{Data assignment matrix for $K=20$ and $\mathcal{I}=\left\{1,4,11,15,6,18\right\}$.}
\label{assign_cc}
\end{figure}
\begin{figure}
 \[
  \begin{bmatrix}
  {\color{blue}\mathbf{g}_{1}} \\
  {\color{red}\mathbf{g}_{4}} \\
  {\color{red}\mathbf{g}_{11}} \\
  {\color{brown}\mathbf{g}_{15}} \\
  {\color{brown}\mathbf{g}_{6}} \\
  {\color{brown}\mathbf{g}_{18}} \\
  \end{bmatrix} \rightarrow
  \begin{bmatrix}
   \widetilde{\mathbf{g}}_{1,1} \\
   \widetilde{\mathbf{g}}_{1,2}  \\
  \widetilde{\mathbf{g}}_{1,3} \\
  \end{bmatrix} =
  \begin{bmatrix}
   \mathbf{g}_{1} \\
   \mathbf{g}_{4} + \mathbf{g}_{11} \\
   \mathbf{g}_{15} + \mathbf{g}_{6} + \mathbf{g}_{18}\\
  \end{bmatrix}.
  \]
  \caption{Illustration of the encoding phase at the first worker for coded communication with the computation assignment matrix in Fig. \ref{assign_cc} and a degree vector $\mathbf{d}=[1,2,3]$.}
 \label{rcsc_encode_2}
\end{figure}

The GC scheme is designed for exact recovery of the summation, and limits the number of messages per worker to one. The MMC variation of  GC is studied before in \cite{UCCT.4,UCCT.6}. Here, we will show that the RCS code proposed for matrix-vector multiplication can also be used for partial recovery in coded communication with a small variation in the encoding phase.\\
\indent We will illustrate RCS coded communication on an example. Consider the previous example with $K=20$ workers. As before, an $L\times K$ computation assignment is formed to assign partial computations to workers with a certain computation order as illustrated in Fig. \ref{assign_cc}. In the coded communication scenario encoding takes place after the computation, therefore the computation load of the computation assignment matrix in Fig. \ref{assign_cc} is $r=L=6$, whereas the same matrix would have a computation load of $r=3$ in the coded computation scenario. Again, once the assignment matrix is formed, coded messages for each worker are constructed according to the given degree vector $\mathbf{d}$  and based on the assignment matrix $\mathbf{A}_{RCS}$ as illustrated in Fig. \ref{rcsc_encode_2}. Note that, similarly to the coded communication scenario RCS code allows recovery of only a subset of the partial computations, and can compute an approximation to the required summation. Note, however, that, while in coded computation missing results correspond to entries of the vector we would like to compute, here the missing results will impact every entry of the desired computation as we will be missing some of the the partial computations. We want to remark that, concurrently to our work, GC scheme with particular focus on the trade-off between computation accuracy and time has been studied in \cite{CC.AG1,CC.AG2,CC.AG3,CCP.1,UCCT.7}.
\begin{figure*}
 \[
  \mathbf{A}_{RCS}=
  \begin{bmatrix}
  {\color{red}\mathbf{W}_{5}} & {\color{red}\mathbf{W}_{6}} & {\color{red}\mathbf{W}_{7}} &  {\color{red}\mathbf{W}_{8}} \\
  {\color{blue}\mathbf{W}_{1}} & {\color{blue}\mathbf{W}_{2}} & {\color{blue}\mathbf{W}_{3}} &  {\color{blue}\mathbf{W}_{4}} \\
  {\color{blue}\mathbf{W}_{3}}\ & {\color{blue}\mathbf{W}_{4}} & {\color{blue}\mathbf{W}_{1}} &  {\color{blue}\mathbf{W}_{2}} \\
  {\color{red}\mathbf{W}_{8}}\ & {\color{red}\mathbf{W}_{5}} & {\color{red}\mathbf{W}_{6}} &  {\color{red}\mathbf{W}_{7}} \\
  {\color{red}\mathbf{W}_{7}} & {\color{red}\mathbf{W}_{8}} & {\color{red}\mathbf{W}_{5}} &  {\color{red}\mathbf{W}_{6}}\\
  \end{bmatrix}
  \]
\caption{$\mathbf{A}_{RCS}$ based on   $\mathbf{z}=[2,1,1,2,2]$, and random samples $\left\{1,3\right\}\in\mathcal{I}_1$, $\left\{1,4,3\right\}\in\mathcal{I}_2$}.
\label{RCS_gen1}
\end{figure*}
 \begin{figure*}
 \[ 
  \mathbf{C}_{RCS}=
  \begin{bmatrix}
  {\color{red}\mathbf{W}_{5}} & {\color{red}\mathbf{W}_{6}} & {\color{red}\mathbf{W}_{7}} &  {\color{red}\mathbf{W}_{8}} \\
  {\color{blue}\mathbf{W}_{1}} & {\color{blue}\mathbf{W}_{2}} & {\color{blue}\mathbf{W}_{3}} &  {\color{blue}\mathbf{W}_{4}} \\
  {\color{blue}\mathbf{W}_{3}} + {\color{red}\mathbf{W}_{8}} +{\color{red}\mathbf{W}_{7}} & {\color{blue}\mathbf{W}_{4}} +{\color{red}\mathbf{W}_{5}} + {\color{red}\mathbf{W}_{8}} & {\color{blue}\mathbf{W}_{1}} + {\color{red}\mathbf{W}_{6}} + {\color{red}\mathbf{W}_{5}} &  {\color{blue}\mathbf{W}_{2}} + {\color{red}\mathbf{W}_{7}} + {\color{red}\mathbf{W}_{6}}\\
  \end{bmatrix}.
  \]
\caption{$\mathbf{C}_{RCS}$ based on $\mathbf{A}_{RCS}$ and $\mathbf{d}=[1,1,3]$}
\label{RCS_gen2}
\end{figure*}

\section{Generalized RCS codes}
In the introduced RCS code structure, the main computation task is divided into $K$ equal sub-tasks. However, if the variation on the computational speed of the workers is small, it might be better to divide them into even smaller tasks in order to better utilize the computational resources\\
\indent Here, we present generalized RCS codes that allow adjusting the sizes of individual computation tasks. In generalized RCS codes, the encoding part remains the same as before, but the construction of the  $\mathbf{A}_{RCS}$ matrix is as follows. First, $\mathbf{W}$ is  divided into $KN$ disjoint submatrices, i.e., $\mathbf{W}_{1},\ldots,\mathbf{W}_{KN}$,  which are then divided into $N$ groups $\overline{\mathbf{W}}^{(1)},\ldots,\overline{\mathbf{W}}^{(N)}$, each containing $K$ submatrices, i.e.,  $\overline{\mathbf{W}}^{(i)}=[\mathbf{W}_{(i-1)K+1},\ldots,\mathbf{W}_{iK}]$. Before the construction of $\mathbf{A}_{RCS}$ matrix, we will define a vector $\mathbf{z}$ which will be useful. \\
\indent Let $\mathbf{z}$ is a $L=\sum^{m}_{i=1}\mathbf{m}(i)$ dimensional vector where each entry is from the set $[N]$. Construction of $\mathbf{A}_{RCS}$ is executed row by row such that for the $i$th row, we first check $z(i)$ and accordingly use the submatricies  in group $\overline{\mathbf{W}}^{(Z(i))}$. Once we decide on the group of submatricies $\overline{\mathbf{W}}^{(Z(i))}$, we randomly sample an element $j$ from set $\in\mathcal{I}_{z(i)}$ and circularly shift the $\overline{\mathbf{W}}^{(Z(i))}$ with $j-1$ before assigning it as the $i$th row of $\mathbf{A}_{RCS}$. We remark that, initially, $\mathcal{I}_{i}=[K]$, $\forall i\in[N]$, and after the circularshift operation based on sampled $j$ we remove the $j$ from the corresponding set $\mathcal{I}_{i}$ to prevent repetition among the rows. The detailed procedure is illustrated in Algorithm \ref{alg_gen_rcs}.
We present a simple example for $K=4$ and $N=2$  to clarify the overall procedure. Let $\mathcal{I}_1=\left\{1,3\right\}$, $\mathcal{I}_2=\left\{1,4,3\right\}$,  and $\mathbf{z}=[2,1,1,2,2]$, and the construction procedure of $\mathbf{A}_{RCS}$ is illustrated in Fig. \ref{RCS_gen1}. The computation assignment matrix $\mathbf{C}_{RCS}$ is given for $\mathbf{d}=[1,1,3]$ in Fig. \ref{RCS_gen2}. As in the previous example illustrated in Fig. \ref{rcsc}, each user is allowed to send at most 3 messages, but now the computation load is $r=3/2$ instead of $3$ as each of the computations is half the size of those in the previous example. If the computation speeds of the workers are more likely to be similar, it will be better to divide the main computation task into smaller subtasks to utilize the available computation resources better. Here, we remark that the first row of $\mathbf{C}_{RCS}$ is from $\overline{\mathbf{W}}^{(2)}$ while the second row is from $\overline{\mathbf{W}}^{(1)}$. Hence, considering only the first two assigned computations, the  recovery probability of $\mathbf{W}_{i}\boldsymbol{\theta}$ is higher for $\mathbf{W}_{i}\in\overline{\mathbf{W}}^{(2)}$, compared to $\mathbf{W}_{i}\in\overline{\mathbf{W}}^{(1)}$. Therefore, the third row is generated using two submatrices  from $\overline{\mathbf{W}}^{(2)}$ and one submatrix from $\overline{\mathbf{W}}^{(1)}$. Consequently, by playing with $\mathbf{d}=[1,1,3]$ and  $\mathbf{z}=[2,1,1,2,2]$ different operating points in terms of the computation speed and accuracy can be achieved.

\begin{algorithm}
\caption{ Generalized RCS coded
computation}\label{alg_gen_rcs}
\begin{algorithmic}[1]
\State{\textbf{Data assignment phase:}}
\State{$L=\sum^{m}_{i=1}d_{i}$}
\For{$j=1:N$}
\State{$\mathcal{I}_j=[K]$}
\EndFor
\For{row index $i=1,2,\ldots,L$}
\State{Randomly choose  $j\in\mathcal{I}_{z(i)}$}
\State{Update $\mathcal{I}_{z(i)}$: $\mathcal{I}_{z(i)} \leftarrow \mathcal{I}_{z(i)} \setminus \left\{ j\right\}$}
\State{$\mathbf{A}(i,:) = circshift(\overline{\mathbf{W}}^{(z(i))},j-1)$}
\EndFor
\State{\textbf{Data encoding phase:}}
\For{worker $k=1,2,\ldots,K$}
\For{message $j=1,\ldots,r$}
\State{Starting row index: $l_s = \sum^{j-1}_{i=1}d_i+1$}
\State{Ending row index: $l_e = \sum^{j}_{i=1}d_i$}
\State{$\widetilde{\mathbf{W}}_{k,j}=\sum^{l_e}_{l=l_s} \mathbf{A}(l,k)$}
\EndFor
\EndFor
\end{algorithmic}
\end{algorithm}

\section{Connection to LT Codes and Choosing the Degree Distribution}

In this section, we will discuss the connections of the proposed RCS coding strategy to LT codes to better understand how the degree distribution can be optimized. First, we briefly summarize the way  decoding process is analyzed in the LT code design \cite{lt, lt2}. In LT codes, the data we want to recover is divided in to $K$ parts, each of them referred to as an {\em input symbol}. The {\em output symbols} are generated as XORed combinations of the $d$ randomly chosen input symbols, where $d$ is a random variable from a discrete distribution $\Omega(d)$. Input symbols XORed to obtain an output symbol are called its {\em neighbors}. LT process consists of three key steps;
\begin{itemize}
    \item (release) When an output symbol of degree one appears, we refer to it as a release.
    \item (cover) Following release of an output symbol, the corresponding unique input symbol is covered.
    \item (process) A covered input symbol is removed from all its neighboring output symbols.
\end{itemize}
All the covered but unprocessed symbols are stored in a buffer called \textit{ripple}. During the flow of the LT process, initially all degree-one codewords are immediately released and stored in the ripple, then at each step of the process, one symbol is chosen to be processed until the ripple size, denoted by $R$, becomes zero, at which point the algorithm halts.\\
\indent The evolution of the ripple size throughout the decoding iterations is the main focus of the LT code design when defining the degree distribution, and the aim is to keep the ripple size as constant as possible to minimize the overhead. Ideal strategy is to have $R=1$ thoughout the LT process \cite{lt}, however, in practice, due to randomness ripple can be emptied before the $K$ input symbols are covered. Hence, in \cite{lt}, {\em robust soliton distribution} is introduced for choosing the degree of the output symbols so that LT process starts with ripple size $R>1$ and kept constant as much as possible.\\
\indent In the next two subsections, we first discuss the impact of the straggling behavior; that is, the statistics of the codewords with particular order, on the design of the degree distribution in our scenario.

\subsection{The impact of the straggling behaviour}
In the scope of this work, to model the statistics of the computational latency we will employ the shifted exponential distribution, similar to the existing works such as \cite{CC.1}, which is defined with two parameters, the mean of the exponential distribution $1/\mu$ and the shift parameter $\alpha$ such that the probability of completing $s$ computations at any server, e.g., performing $s$ identical  matrix vector multiplications, by time $t$ is given by
\begin{equation}\label{dist}
F_{s}(t)\defeq
    \begin{cases}
     1-e^{-\mu(\frac{t}{s}-\alpha)}, &  \text{if } t\geq s\alpha,\\
      0,   &  \text{otherwise}. 
    \end{cases}
\end{equation}
We emphasize that since the $\alpha$ parameter simply denotes the minimum required time to complete a single computation and as we increase it, while keeping $\mu$ constant, computation latency of the workers become more and more correlated. Consequently, if we use $0\leq\rho_{i}\leq1$ to denote the fraction of the codewords with order $i$ collected at the PS, then $\rho_{1}$ increases in parallel to increase in the shift parameter. We remark that, although this analysis is not limited to shifted exponential distribution we particularly use it here, since it clearly reflects the impact of the straggling behaviour on the code design. Now, we consider a simulation setup where we assume $K=40$ and $r=3$ and use the shifted exponential model for the computational statistics with $\mu=10$ and 4 different $\alpha$ values, $0.01,0.05,0.1$ and $0.2$, respectively. In Table \ref{fraction}, we illustrate the fraction the codewords with a particular order within the first $K$ received messages. At this point we want to highlight that the major alteration compared to LT code is to utilize the side information regarding the erasure probability in the design of the degree distribution. To be more precise, in robust solution distribution, probability of degree one codewords kept small in order to prevent reception of duplicate symbol, however RCS further utilizes the degree one codewords since degree one codewords from different worker do not overlap. To clarify let $\tilde{\Omega}_{(K)}(d)$ denote degree distribution observed at PS when $K$ messages are received in total. In the case of RCS and shifted exponential distribution with parameters $\mu=10$ and $\alpha=0.01$ we observe $\tilde{\Omega}_{(K)}(1)=0.55$. At this point we want to highlight the following fact that with RCS there is no overlapping degree one codewords, however, using the same distribution for the LT codes induce on average $K/10$ number of overlapping degree one codewords which implies around \%10 redundancy. The use of degree one codewords is one of the major difference between the LT code design  and the RCS.

\begin{table}[t]
\center
\begin{tabular}{|l|l|l|l|}
\hline
\multirow{2}{*}{Shift parameter} & \multicolumn{3}{l|}{Order of the Codewords} \\ \cline{2-4} 
                               &     $\rho_{1}$              &     $\rho_{2}$             &       $\rho_{3}$            \\ \hline
$\alpha=0.01$                           & 0.55                & 0.3                & 0.15              \\ 
\hline
$\alpha=0.05$                            & 0.69                & 0.2750           & 0.0375                \\ 
\hline
$\alpha=0.1$                             & 0.78                & 0.22          & 0                \\ 
\hline
$\alpha=0.2$                            & 0.9               & 0.1             & 0               \\ \hline

\end{tabular}
\caption{Fraction of the codewords with a particular order with respect to different shift parameters, $\alpha$.}
\label{fraction}
\end{table}

\subsection{Incorporating robust soliton distribution}
\begin{figure*}
    \centering
         \begin{subfigure}[b]{0.45\textwidth}
        \includegraphics[scale=0.6]{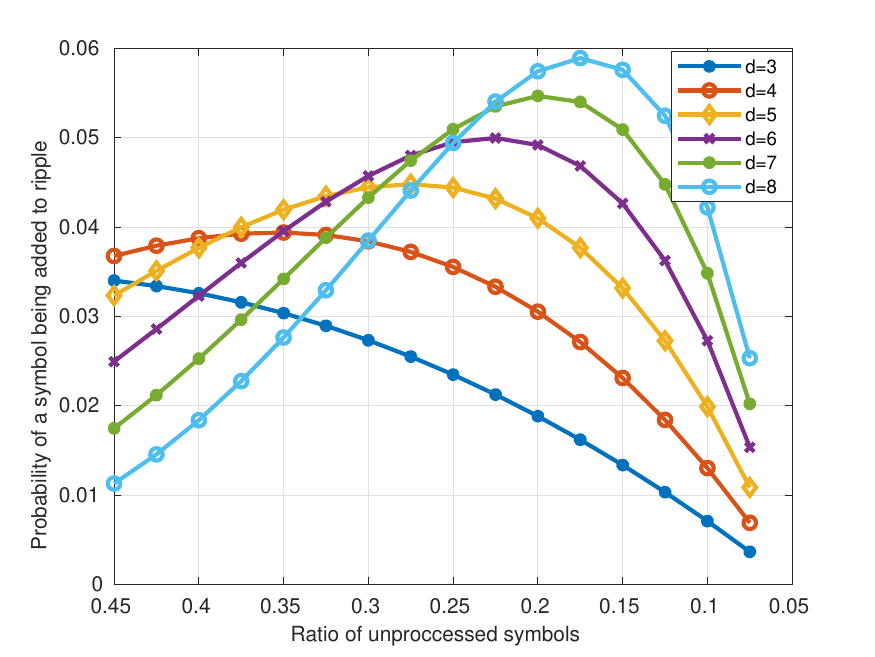}
        \caption{$\rho_{rip}=0.05$.}
    \end{subfigure}
    \begin{subfigure}[b]{0.45\textwidth}
        \includegraphics[scale=0.6]{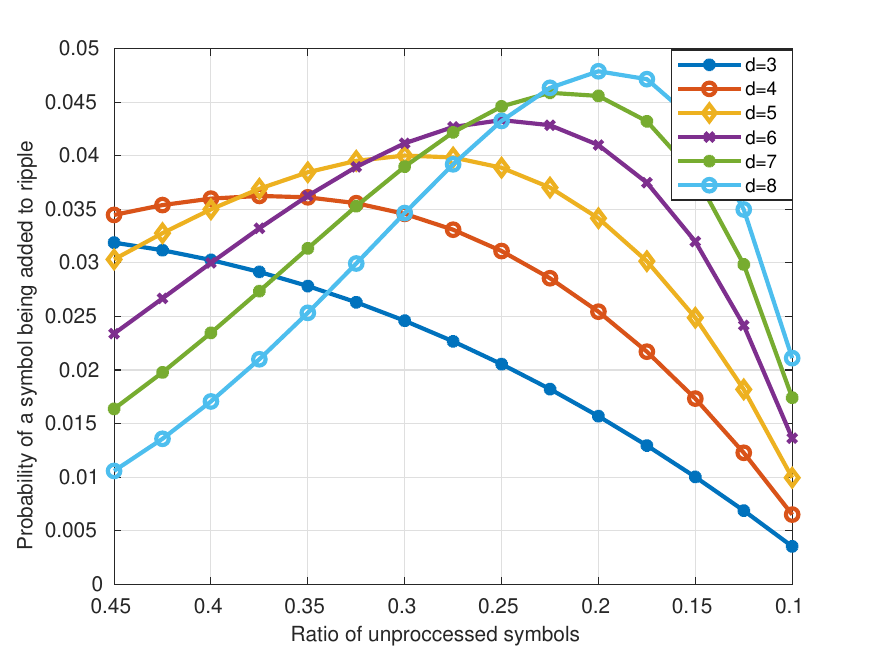}
        \caption{$\rho_{rip}=0.075$}
        \end{subfigure}
    \begin{subfigure}[b]{0.45\textwidth}
        \includegraphics[scale=0.6]{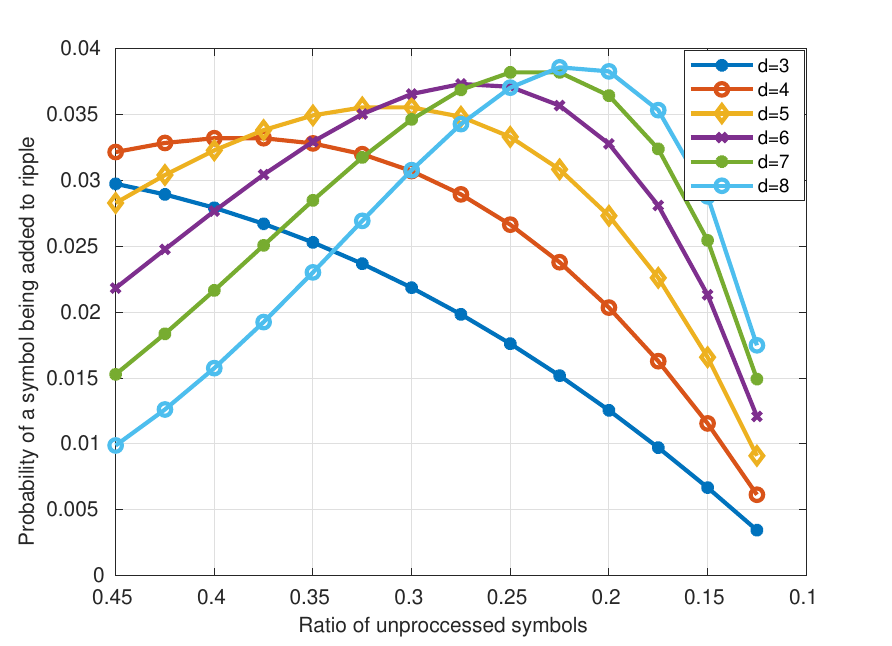}
        \caption{$\rho_{rip}=0.1$.}
    \end{subfigure}
    \caption{The impact of the codeword degree on the ripple evaluation with respect to the initial ripple ratio $\rho_{rip}$ for $\rho=0.55$.}
		\label{releasefig}
\end{figure*}

To provide a general perspective on the impact of the degree distribution, we first focus on one-shot recovery such that the recovery of a symbol from a received codeword of degree $d$, given that $\rho$ portion of the symbols are already recovered. Based on the results illustrated in Table \ref{fraction}, we are particularly interested in a regime where  $\rho>0.5$.\\
\indent Given values of $K$, $\rho$ and $d$, probability of recovering an input symbol from a codeword/output symbol of degree $d$ is
\begin{equation}
\frac{{K\rho\choose d-1}{K-K\rho\choose 1}}{{K \choose d}} =\Pi^{d-2}_{i=1}\frac{K\rho-i}{K-i}\frac{d(K-K\rho)}{K-d-1},
\label{decodeprob}
\end{equation}
further if $K$ is sufficiently large,  also assuming $K\rho>>d$, this can be approximated as
\begin{equation}
\rho^{d-1}d(1-\rho).
\label{redundancyprob}
\end{equation}
Hence, for a given $\rho$, one can choose  $d$ that maximizes the (\ref{decodeprob}) as a greedy approach. This can be considered as the optimal degree choice under the assumption that codeword is discarded immediately if an input symbol can not be recovered. However, in LT process codewords are buffered and might be used to recover an input symbol at any step of the decoding phase. Hence, we consider $d$ that maximizes maximizes the (\ref{decodeprob}) as a lower bound and denoted with $d_{min}$. One can easily observe that $d_{min}$ increases with respect to $\rho$, e.g., for $R$ values around $0.5$, $0.6$, $0.7$ and $0.8$, $d_{min}$ takes values around $2,3,4$ and $5$, respectively. Hence, we can emphasize that there is no {\em once-for-all} solution and the degree distribution should be designed based on the computation statistics. The reason  we define $d_{min}$ as the lower bound for the codeword degree is that, although it maximize the probability of recovering a new input symbol in a one-shot manner, it also give rise to redundancy  that is the probability of a codeword having all its neighbours from the $R$ recovered input symbols. This probability can be approximated as $\rho^{d}$.\\
\indent To understand the dynamic nature of the decoding process we revisit the analysis of the robust soliton distribution  \cite{lt}. The objective of the robust soliton distribution is to initially obtain a small but sufficient ripple size $R$ from degree one codewords and choose the degree distribution for $\Omega(d)$ for $d>2$ in a way to keep ripple size constant through the decoding process. Hence, to keep the ripple size constant the number of released symbol should be kept around one, on average. Given that the ripple size is $R$ and the number of unprocessed symbols is $L$, the probability of a codeword  of degree $d$ being released and an input symbol being added to ripple is equal to
\begin{align}\label{release}
q(d,L,R)=&\frac{{K-(L+1)\choose d-2}{L-R\choose 1}}{{K \choose d}}\\
=& \frac{d(d-1)(L-R)\Pi^{d-3}_{i=0} (K-(L+1)-i)}{\Pi^{d-1}_{i=0}(K-i)}.
\end{align}
Hence, robust soliton distribution aims to keep density of the codewords with degree $d$ proportional to $\frac{1}{d(d-1)(L-R)}$ so that at each step, on average, one symbol is released and added to ripple. As we already aforementioned, the key difference between the RCS and the LT code design is the density of the degree one codewords, which is also the reason why it is not possible to follow the same logic used in the robust soliton distribution in this setup. To be more precise, the RCS strategy implies a large initial ripple size $R$ compared to LT codes. Here, we remark that when a symbol is released, probability of released symbol being added to ripple is $L-R/L$, hence if one try to keep $R$ constant, then large portion of the  released symbols will be redundant.\\
\indent Hence, we slightly change the analysis of the robust soliton distribution in the following way; first, as in the original analysis, all degree one codewords are added to the ripple immediately, then instead of one-by-one processing, we process all $R=\rho K$ symbols at the same time, thus $R$ is reduced to $0$. Then, new released symbols, following the process of $R=\rho K$ symbols, are added to ripple. At this point we use the equation (\ref{decodeprob}) to compute estimate of the new ripple size. Then, we follow analysis of robust soliton distribution by processing exactly one symbol at each time and try to keep $R$ as much as constant. By achieving new $R$ value that is comparatively smaller than $\rho K$  the redundancy is avoided.\\
\indent In case of robust soliton distribution the ripple size $R$ is only depend on the density of degree one output symbols, but now in our case $R$ is actually depends on the whole degree distribution $\Omega(d)$ such that $R$ will be  bounded by 
\begin{align}\label{ripple_size}
R\leq &K(1-\rho)\sum^{d_{max}}_{d=2}\Omega(d)\rho^{d-1}d(1-\rho)\\
\rho_{rip}\leq &\underbrace{(1-\rho)^{2}\sum^{d_{max}}_{d=2}\Omega(d)\rho^{d-1}d}_{\tilde{\rho}_{rip}},\nonumber
\end{align}
where $\rho_{rip}$ is the ratio of the symbols at the ripple. Although (\ref{ripple_size}) gives an upper bound on $\rho_{rip}$, this upper bound is tight, $\rho_{rip}\approx\tilde{\rho}_{rip}$, when $\tilde{\rho}_{rip}<<1-\rho$ and $K$ sufficiently large. This is equivalent to observing that when $R$ balls are chosen out of $L$ with replacement, the number of distinct balls are approximately equal to $R$ for $R/L$ is sufficiently small. Therefore, we ignore the redundancy within the released symbols, but focus on the redundant output symbols whose all neighbours are among the initially released $\rho K$ input symbols. The ratio of such output symbols can be approximated as
\begin{equation}
\rho_{red}=(1-\rho)\sum^{d_{max}}_{d=2}\Omega(d)\rho^{d}.
\end{equation}
In the case of full recovery this implies that at least $K\rho_{red}$ extra codewords are required. To exemplify numerically, when $\rho=0.55$, assuming all the remaining codewords are of degree two gives $\rho_{red}=0.135$, and assuming degree three gives $\rho_{red}=0.0745$. Hence, although robust soliton distribution has a peak around $d=2$, in our case we expect $\Omega(d)$ has a peak at a higher $d$ values to avoid redundancy. In overall, there are three critical aspects when deciding the degree distribution, particularly to target the full recovery, keeping $\rho_{red}$ as close to zero as possible, having small but sufficient ripple ratio $\rho_{rip}$ and keeping $\sum^{d_{max}}_{d=2}\Omega(d)q(d,L,R)$ close to $\frac{1}{K\rho}$ in order to preserve the ripple size. In Fig. \ref{releasefig}, we illustrate the probability of symbol being released from a codeword of degree $d$ and added to the ripple for values $\rho_{rip}$ of $0.05,0.075$, and $0.1$, respectively. \\
\indent Hence, if we revisit our particular scenario where $r=3$ and $\alpha=0.01$ and target the full recovery then, based on the aforementioned key aspects, a proper choice for degree distribution would be $\mathbf{d}=[1~4~ 8]$ or $\mathbf{d}=[1~5~7]$, also taking into account that $\rho_{2}=0.3$ and $\rho_{3}=0.15$. Once,  the best possible degree distribution for the full recovery is available, then based on the requirements, one can search for the appropriate degree distribution by reducing degree values while still trying to keep $\rho_{rip}$ as small as possible. In Table \ref{trade-off}, we illustrate this trade-off by comparing the total number of messages received until the termination with a tolerance $q$.

\begin{table*}[h]
\center
\begin{tabular}{|l|l|l|l|}
\hline
\multirow{2}{*}{Degree vector} & \multicolumn{3}{l|}{Average Number of Codewords} \\ \cline{2-4} 
                               &     $q=0.3$              &     $q=0.15$             &       $q=0.0$            \\ \hline
$\mathbf{d}=[1~2~3]$                           & 34                & 42.7                 & 64.3                \\ \hline
$\mathbf{d}=[1~3~3]$                           & 34.9                & 41.2                 & 58.6              \\ \hline
$\mathbf{d}=[1~3~4]$                              & 36                & 41.6           & 55.6                \\ \hline
$\mathbf{d}=[1~3~5]$                              & 37               & 41.6           & 53.6                 \\ \hline
$\mathbf{d}=[1~4~4]$                              & 37.06               & 41.17           & 52.1                \\ \hline
$\mathbf{d}=[1~5~7]$                             & 41.7               & 43.8              & 46.6               \\ \hline
$\mathbf{d}=[1~4~8]$                              & 40.9                 & 44.3              & 47.24             \\ \hline
\end{tabular}
\caption{Average number of required codewords to complete an iteration with a tolerance of $q$ }
\label{trade-off}
\end{table*}

\subsection{Changing the decoding strategy for a better trade-off}

In the previous subsection, we highlighted the main aspects on the design of degree distribution when the LT process is utilized for decoding. However, it might be possible to achieve better trade-offs by altering the decoding strategy as partly discussed in \cite{partrec}. Gaussian elimination method can be used instead of LT process, but in general it is a computationally costly procedure. Nevertheless, a hybrid approach can be utilized to achieve better decoding performance without excessive computational overhead. That is, at the beginning LT process is  followed to recover input symbols and when it halts Gaussian elimination is applied to remaining codewords. The key advantage of such strategy will be  reducing the redundancy. Next, we illustrate this trade-off with an example in our setup. 
\begin{table*}[h]
\center
\begin{tabular}{|l|l|l|l|}
\hline
\multirow{2}{*}{Degree vector and decoding process} & \multicolumn{3}{l|}{Average Number of Codewords} \\ \cline{2-4} 
                               &     $q=0.3$              &     $q=0.15$             &       $q=0.0$            \\ \hline
$\mathbf{d}=[1~5~7]$ Hybrid                          & 38.4                & 39.9                 & 44.4                \\ \hline
$\mathbf{d}=[1~5~7]$ LT process                              & 41.7                & 43.8           & 46.65                 \\ \hline
$\mathbf{d}=[1~3~5]$ Hybrid                          & 35.8               & 40.85                 & 53               \\ \hline
$\mathbf{d}=[1~3~5]$ LT process                              & 37                & 41.6          & 53.6                \\ \hline
\end{tabular}
\caption{Average number of required codewords to complete an iteration with a tolerance of $q$ }
\label{hybrid}
\end{table*}
As demonstrated in Table \ref{hybrid}, change of the decoding process offers better trades-offs, especially when higher degree codewords, e.g., $\mathbf{d}=[1~5~7]$, are used when it is desired to reduce the number of received messages at the operating point $q=0$, but still want to achieve lower latency at $q=0.3$ and $q=0.1$. In the case of LT process, to achieve similar performance at the regime of $q=0.3$ and $q=0.15$, it is required to reduce degree of codewords, such as using $\mathbf{d}=[1~ 3 ~5]$, which induce a visible overhead for the full recovery in terms of the number of received messages.

\section{Numerical Results and Discussions}\label{simulations}
For the numerical analysis, we will first analyze the convergence performance of the partial recovery strategy for coded computation with RCS codes under different tolerance requirements, then we will compare the \textit{average per-iteration completion time}  of the RCS code with UC-MMC and MDS coded computation (MCC) schemes, and, finally we will extend our analysis to coded communication.

\subsection{Simulation setup}
For the simulations, we consider a linear regression problem over synthetically created training  dataset, according to normal mixture distribution as in \cite{CC.4},  consisting of size  of 2000 samples. In the generation of synthetic data, we first create the true parameter model $\mathbf{\theta}^{\star}$ randomly sampling each entry from the interval $[0,1]$ according to uniform distribution. Once we have the $\mathbf{\theta}^{\star}$, we construct two mean vectors $\mathbf{\mu}_1 = \frac{1.5}{d}\mathbf{\theta}^{\star}$ and $\mathbf{\mu}_2= \frac{-1.5}{d}\mathbf{\theta}^{\star}$, where this mean vectors are later used in the normal mixture distribution $\frac{1}{2}\mathcal{N}(\mathbf{\mu}_1,\mathbf{I}) +\frac{1}{2}\mathcal{N}(\mathbf{\mu}_2,\mathbf{I})$ to generate the each row of dataset $\mathbf{X}$.\\
\indent We assume that there are  $K=40$ homogeneous workers, and set $\mu=10$ and $\alpha=0.01$ for the statistics of their computation speeds in (\ref{dist}). For RCS coded computation we choose the degree vector $\mathbf{d}=[1, 2, 3]$, which corresponds to computation load of $r=3$, and we execute  Algorithm 1 accordingly. In all the simulations, we set the  learning rate to $\lambda=0.1$.

\subsection{Simulation results}
\begin{figure*}
\centering
\includegraphics[scale=0.3]{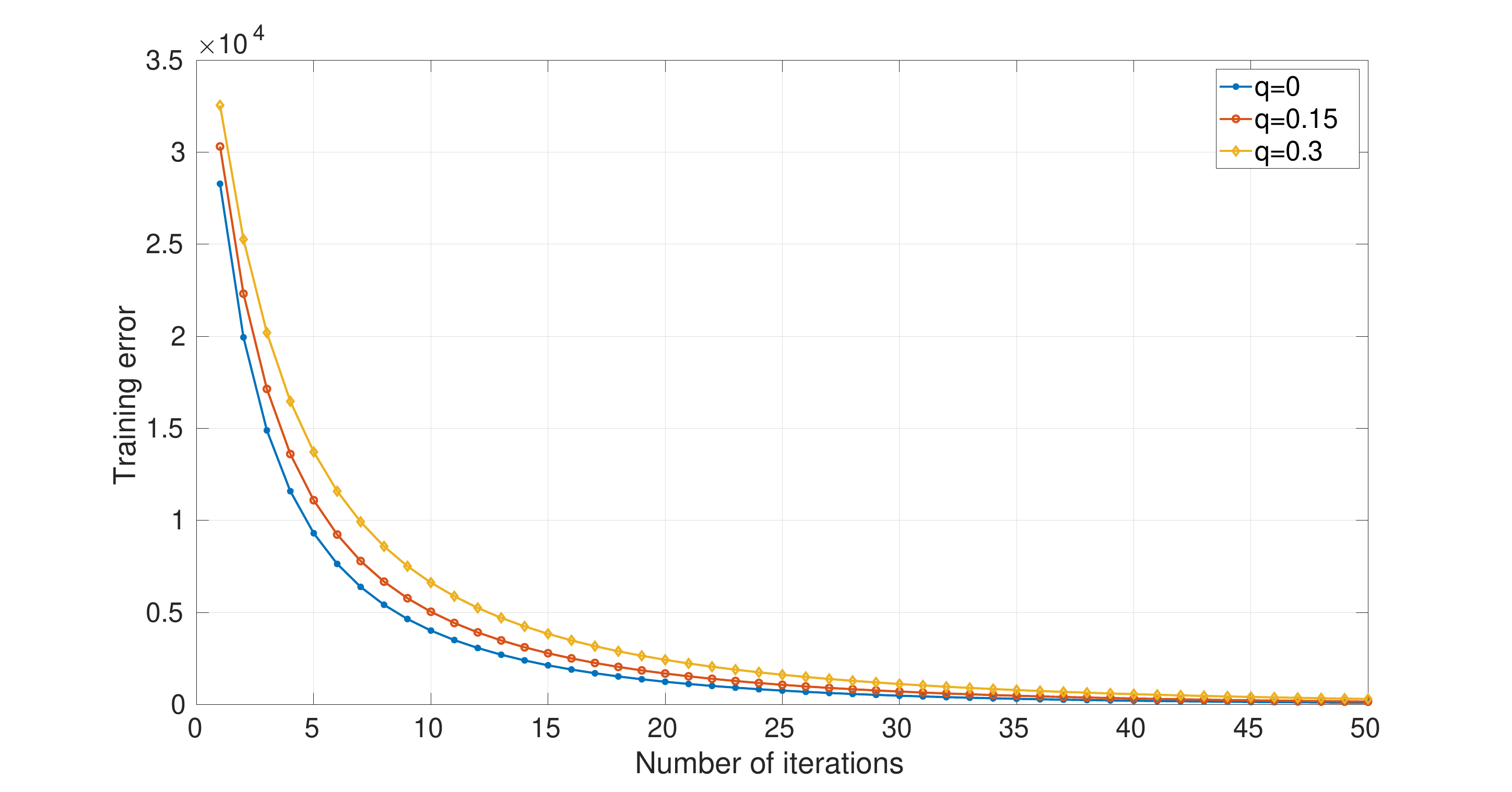}
\caption{Training error over $T=50$ iterations, for a model size of $d=800$, and tolerance level of $q=0,0.15,0.3$, respectively.}
\label{d800}
\end{figure*}

\begin{figure*}
\centering
\includegraphics[scale=0.3]{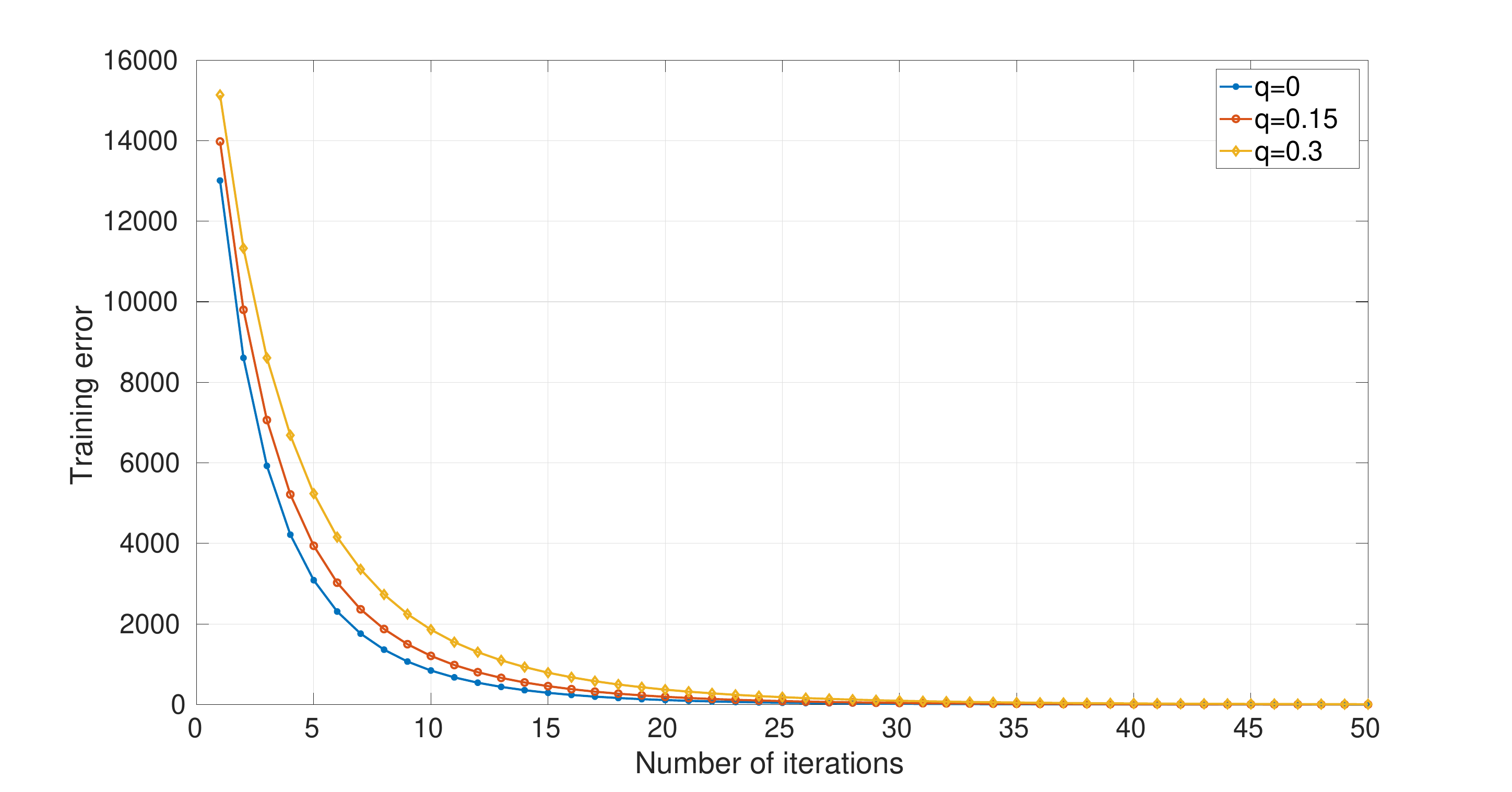}
\caption{Training error over $T=50$ iterations, for a  model size of $d=400$, and tolerance level of $q=0,0.15,0.3$, respectively}
\label{d400}
\end{figure*}

\indent We first consider a model size of $d=800$, and evaluate the training error over $T=50$ iterations\footnote{ For the convergence plot, we take average over 100 independent simulations.} for tolerance level of $q=0$ (which corresponds to full recovery), $q=0.15$, and $q=0.3$. One can observe in Fig. \ref{d800}  that, although the convergence speed reduces with increasing tolerance level at each iteration, partial recovery does not harm the convergence behaviour much, especially if the tolerance level is moderate, e.g., $q=0.15$. We then repeat the same experiments for a model size of $d=400$, which demonstrates similar trends as seen in Fig. \ref{d400}.\\
\indent What we  want to see next is how much reduction in per-iteration time can be achieved by the partial recovery scheme.

\begin{table*}[]
\centering
\begin{tabular}{|l|l|l|l|l|l|l|l|}
\hline
\multirow{2}{*}{Computation strategy / tolerance } & \multicolumn{3}{l|}{RCS} & \multicolumn{3}{l|}{UC-MMC} & \multirow{2}{*}{MCC} \\ \cline{2-7}
                                      & $q=0.0$ & $q=0.15$  & $q=0.3$  & $q=0$    & $q=0.15$    & $q=0.3$   &                      \\ \hline
Average iteration time                &  0.1475      &  0.0936       & 0.0776       &   0.2424     &   0.1170        &   0.0799      &    0.1572                  \\ \hline
Number of received messages              &  60.93    &   42.38     & 35.03      &   81.29    &  51.16       &   36.70     &    14                \\ \hline

\end{tabular}
 \caption{Comparison of the proposed RCS scheme with UC-MMC and MCC schemes for the computational load $r=3$}
 \label{tab:linear}
\end{table*}
Hence, we present the per-iteration time of three different schemes, namely RCS, UC-MMC and MCC, for computational load $r=3$. For RCS, we use the order vector $\mathbf{m}=[1,2,4]$, for UC-MMC we use cyclic shifted assignment as in \cite{CC.5}, and finally, for MCC we use $(\lceil K/r \rceil, K) = (14, 40)$ MDS code.\\
\indent We compare the three schemes in terms of two performance metrics: {\em average completion time} and {\em number of received messages}, which demonstrate how fast an iteration will be completed, and the induced communication load, respectively. From Table \ref{tab:linear}, one can observe that for  full recovery, i.e., $q=0$, RCS outperforms  both  UC-MMC and MCC schemes. We also observe that by allowing partial recovery  it is possible to achieve approximately $40\%$ to $50\%$ reduction in the per iteration completion time with $q=0.15$ and  $q=0.30$, respectively. Here, we note that the partial recovery approach can be also employed with UC-MMC; however, as demonstrated in Table \ref{tab:linear}, RCS outperforms UC-MMC for all given tolerance values. Besides, with RCS PS completes an iteration with less number of received messages, which means that compared to UC-MMC, RCS induces less communication load and congestion. For instance, for $q=0$ UC-MMC requires, on  average, $28\%$ more messages to complete an iteration compared to the proposed RCS scheme, which is shown to be a critical factor affecting  the performance of  real implementations \cite{overview}. We note that MCC requires the minimum number of messages to complete an iteration. Hence, for $q=0$ MCC can be a better alternative; nevertheless,  another advantage of RCS compared to MCC is that the decoding process can be executed in parallel to computations so that the additional latency due to decoding is minimized. Finally, based on the simulation results, we can conclude that the RCS scheme operates most efficiently when  partial recovery is aimed with a low tolerance level, e.g., $q=0.15$, since in this regime we can see the advantage of both using coded computation and partial recovery. To clarify this point, we observe that increasing the tolerance level to $q=0.3$ makes considerable impact on the training accuracy, but the reduction on the average per-iteration computation time is relatively small. Besides, as $q$ increases the performance of the UC-MMC scheme gets closer to that of RCS.\\
\indent We also conduct an experiment for the generalized RCS code with $N=2$. For the simulation we choose $d=[1, 1, 4, 8]$  so that  4 coded computations are assigned to each worker in total and the complexity of the each computation is half of the original RCS schemes, hence $r=2$. The coded computations contain one submatrix from the first group, one submatrix from the second group, 3 submatrices from each group and 5 submatrices from each group, respectively\footnote{Corresponding vector $\mathbf{z}=[1, 2, 1, 1, 2, 2, 1, 1, 1, 1, 2, 2, 2, 2]$.}. For the given code structure we measure the average iteration time as $0.121$, $0.087$ and $0.075$ for $q=0,0.15,0.3$, respectively. One can observe from the results that by using smaller tasks both computation time and the computation load can be reduced. However, on the other hand use of smaller subtasks may increase the number of messages, for instance, given simulation setup the average number of received messages at the PS are $98$, $80$ and $70$ for $q=0,0.15,0.3$, respectively.

\indent Although the RCS code is initially designed for coded computation, as explained in Section \ref{section:coded_comm}, can be implemented for coded communication as well. Therefore, we repeat our experiments to compare the performances of RCS, UC-MMC and GC for a computational load of $r=6$. For the RCS we now use the degree vector $\mathbf{d}=[1, 2, 3]$. The simulation results are  presented in Table \ref{tab:nonlinear}. We observe that in terms of the average per-iteration completion time UC-MMC and RCS both outperform GC, with UC-MMC typically achieving the lowest computation time. In terms of the communication load, UC-MMC has the worst performance. Hence, the key advantage of RCS in the coded communication scenario is a better balance between the computation  and  communication latencies. At this point, we also want to highlight that UC-MMC can be considered as a special case of RCS, where the degrees of the all message are one. Overall, one can play with the degree vector to achieve different points on the trade-off between the  communication and computation latencies.
\begin{table*}[]
\centering
\begin{tabular}{|l|l|l|l|l|l|l|l|}
\hline
\multirow{2}{*}{Computation strategy / tolerance rate} & \multicolumn{3}{l|}{RCS} & \multicolumn{3}{l|}{UC-MMC} & \multirow{2}{*}{GC} \\ \cline{2-7}
                                      & $q=0.0$ & $q=0.15$  & $q=0.3$  & $q=0$    & $q=0.15$    & $q=0.3$   &                      \\ \hline
Average iteration time                &  0.2219     &  $0.1231$       & 0.0940       &   0.1874     &   0.0986       &   0.0736     &    1.2575                 \\ \hline
Number of received messages              &  $62.56$     &   41.55      & 32.37      &   99.63    &  55.06       &   38.30     &    35                \\ \hline
\end{tabular}
 \caption{Comparison of the proposed RCS scheme with UC-MMC and GC schemes for the computational load $r=6$}
 \label{tab:nonlinear}
\end{table*}

\subsection{Discussions}
We have proposed  a new code construction framework for straggler-aware coded computation/communication schemes, which provides the flexibility to trade-off
the accuracy of computation with the  computation latency and the communication load. While we have provided a specific code construction, several improvements and/or adaptations of this design are possible. Below, we briefly discuss some of the possible future extensions.
\subsubsection{Double threshold scheme}
One of the  possible extensions is to use two thresholds to decide when to terminate an iteration. The reason behind the use of two thresholds strategy is that partial recovery strategy is efficient when the sacrifice from the computation accuracy provides a noticeable reduction in the latency. However, if all the workers are sufficiently fast, then partial recovery may not bring noticeable reduction in the latency but lose accuracy. To this end, after sending the latest model vector $\boldsymbol{\theta}_{t}$, the PS starts keeping time and collects messages from workers for a given fixed duration. Once the duration is completed, PS checks whether the requirement due to tolerance level is satisfied or not and  if it is not satisfied then continues to receive messages from workers until it is satisfied. 
\subsubsection{Adaptive tolerance }
In the case of iterative training of machine learning models, it is known that the update accuracy has different impacts on the convergence at different phases of the training process. Hence, the tolerance  can be adjusted over time to obtain a better overall convergence result. 
\subsubsection{Memory enhanced updates}
Again, when partial computations are used in the context of iterative optimization or training, the PS can benefit from the computations recovered in the previous iterations. In our simulations, at each iteration we use only the computations recovered in that iteration. Instead, it can  utilize the results from previous iterations to compensate for the missing computation results in the current iteration, similar to the {\em momentum SGD} framework.

\section{Conclusions}
In this paper, we have introduced the CCPR approach, and a particular code structure, called RCS, in order to provide an additional flexibility in seeking a balance  between  the per-iteration completion time, the computation accuracy, and the communication load in distributed computing. In particular, for a matrix-vector multiplication task, the RCS code can adaptively recover a portion of the element of the resultant vector. The proposed code construction is built upon the LT code structure, but requires additional optimization of the underlying degree distribution due to the correlation among the erasures of symbols in the code. We have also shown that the RCS code can also provide a similar flexibility in distributed computation of any arbitrary computation task that can be written as the summation of multiple partial computations. We have applied the proposed RCS code to iterative SGD in a linear regression problem. By conducting  experiments using different tolerance values we showed that the RCS code can help to reduce the per-iteration completion time for a reasonable reduction in the update accuracy, which can be tolerated due to the iterative nature of the algorithm. We also showed that, compared to UC-MMC, which can also employ partial recovery, RCS requires, on average, less number of messages to complete an iteration, which means lower communication load. 
\bibliographystyle{IEEEtran}
\bibliography{IEEEabrv,ref.bib}

\begin{IEEEbiographynophoto}{Emre Ozfatura}
 received his B.Sc. in Electronics Engineering with Math minor and M.Sc. in Electronics Engineering from Sabanci University (Turkey), in 2012 and 2015, respectively. He is currently pursuing his Ph.D. degree at Imperial College London, UK, where he is a member of the Information Processing and Communications (IPC) Lab. His research interests are video streaming applications, distributed content storage and distributed computation. 
\end{IEEEbiographynophoto}

\begin{IEEEbiographynophoto} {Sennur Ulukus} (Fellow, IEEE)
is the Anthony Ephremides Professor in Information Sciences and Systems in the Department of Electrical and Computer Engineering at the University of Maryland at College Park, where she also holds a joint appointment with the Institute for Systems Research (ISR). Prior to joining UMD, she was a Senior Technical Staff Member at AT$\&$T Labs-Research. She received the Ph.D. degree in Electrical and Computer Engineering from Wireless Information Network Laboratory (WINLAB), Rutgers University, and the B.S. and M.S. degrees in Electrical and Electronics Engineering from Bilkent University. Her research interests are in information theory, wireless communications, machine learning, signal processing and networks; with recent focus on private information retrieval, age of information, machine learning for wireless, distributed coded computing, group testing, physical layer security, energy harvesting communications, and wireless energy and information transfer.

Dr. Ulukus is a fellow of the IEEE, and a Distinguished Scholar-Teacher of the University of Maryland. She received the 2003 IEEE Marconi Prize Paper Award in Wireless Communications, the 2019 IEEE Communications Society Best Tutorial Paper Award, the 2020 IEEE Communications Society Women in Communications Engineering (WICE) Outstanding Achievement Award, the 2020 IEEE Communications Society Technical Committee on Green Communications and Computing (TCGCC) Distinguished Technical Achievement Recognition Award, a 2005 NSF CAREER Award, the 2011 ISR Outstanding Systems Engineering Faculty Award, and the 2012 ECE George Corcoran Outstanding Teaching Award. She was a Distinguished Lecturer of the IEEE Information Theory Society for 2018-2019.

She is an Area Editor for the IEEE Transactions on Wireless Communications (2019-present) and a Senior Editor for the IEEE Transactions on Green Communications and Networking (2020-present). She was an Area Editor for the IEEE Transactions on Green Communications and Networking (2016-2020), an Editor for the IEEE Journal on Selected Areas in Communications-Series on Green Communications and Networking (2015-2016), an Associate Editor for the IEEE Transactions on Information Theory (2007-2010), and an Editor for the IEEE Transactions on Communications (2003-2007). She was a Guest Editor for the IEEE Journal on Selected Areas in Communications (2008, 2015 and 2021), Journal of Communications and Networks (2012), and the IEEE Transactions on Information Theory (2011). She is the TPC chair of 2021 IEEE Globecom, and was a TPC co-chair of 2019 IEEE ITW, 2017 IEEE ISIT, 2016 IEEE Globecom, 2014 IEEE PIMRC, and 2011 IEEE CTW.
\end{IEEEbiographynophoto}

\begin{IEEEbiographynophoto}
{Deniz G{\"u}nd{\"u}z} [S’03-M’08-SM’13] received the B.S. degree in electrical and electronics engineering from METU, Turkey in 2002, and the M.S. and Ph.D. degrees in electrical engineering from NYU Tandon School of Engineering (formerly Polytechnic University) in 2004 and 2007, respectively. After his PhD, he served as a postdoctoral research associate at Princeton University, as a consulting assistant professor at Stanford University, and as a research associate at CTTC in Barcelona, Spain. ln Sep. 2012, he joined the Electrical and Electronic Engineering Department of Imperial College London, UK, where he is currently a Professor of Information Processing, and serves as the deputy head of the Intelligent Systems and Networks Group. He is also a part-time faculty member at the University of Modena and Reggio Emilia, Italy, and has held visiting positions at University of Padova (2018-2020) and Princeton University (2009-2012). 

His research interests lie in the areas of communications and information theory, machine learning, and privacy. Dr. Gündüz is a Distinguished Lecturer for the IEEE Information Theory Society (2020-22). He is an Area Editor for the IEEE Transactions on Information Theory, IEEE Transactions on Communications, and the IEEE Journal on Selected Areas in Communications (JSAC) - Special Series on Machine Learning in Communications and Networks. He also serves as an Editor of the IEEE Transactions on Wireless Communications. He is the recipient of the IEEE Communications Society - Communication Theory Technical Committee (CTTC) Early Achievement Award in 2017, a Starting Grant of the European Research Council (ERC) in 2016, and several best paper awards. 
\end{IEEEbiographynophoto}

\ifCLASSOPTIONcaptionsoff
  \newpage
\fi




\end{document}